\documentclass[aps,prd,10pt,twocolumn,preprintnumbers,superscriptaddress,nofootinbib,letterpaper]{revtex4}

\usepackage{amsmath,amssymb}
\usepackage{slashed}
\usepackage{graphicx}
\usepackage{booktabs}
\usepackage[dvipsnames]{xcolor}
\usepackage{svg}
\usepackage[export]{adjustbox}
\usepackage[caption=false]{subfig}
\usepackage[colorlinks=true,citecolor=blue,linkcolor=blue]{hyperref}
\usepackage[capitalize]{cleveref}
\crefname{section}{Sec.}{Sec.}

\newcommand{\nocontentsline}[3]{}
\newcommand{\tocless}[2]{\bgroup\let\addcontentsline=\nocontentsline#1{#2}\egroup}

\renewcommand{\vec}[1]{{\mathbf{#1}}}
\newcommand{\orcid}[1]{\href{https://orcid.org/#1}{\includesvg[width=10pt]{plots/orcid}}}
\newcommand{\github}[1]{\href{https://github.com/#1}{\includesvg[width=10pt]{plots/github}}}

\begin{document}

\title{Probing Secret Interactions of Astrophysical Neutrinos in the High-Statistics Era}
         
\author{Ivan Esteban \orcid{0000-0001-5265-2404}\,}           
 \email{esteban.6@osu.edu }
 \affiliation{Center for Cosmology and AstroParticle Physics
  (CCAPP), Ohio State University, Columbus, Ohio 43210, USA}
\affiliation{Department of Physics, Ohio State University, Columbus, Ohio 43210, USA}
\author{Sujata Pandey \orcid{0000-0002-0299-8324}\,}            \email{phd1501151007@iiti.ac.in}
\affiliation{Discipline of Physics, Indian Institute of Technology, \\ Indore, Khandwa Road, Simrol, Indore - 453 552, India}
\author{Vedran Brdar \orcid{0000-0001-7027-5104}}	       \email{vedran.brdar@northwestern.edu}
\affiliation{Fermi National Accelerator Laboratory, Batavia, IL, 60510, USA}
\affiliation{Northwestern University, Department of Physics \& Astronomy,\\ 2145 Sheridan Road, Evanston, IL 60208, USA}
\author{John F. Beacom \orcid{0000-0002-0005-2631}}            
\email{beacom.7@osu.edu}
\affiliation{Center for Cosmology and AstroParticle Physics
  (CCAPP), Ohio State University, Columbus, Ohio 43210, USA}
\affiliation{Department of Physics, Ohio State University, Columbus, Ohio 43210, USA}
\affiliation{Department of Astronomy, Ohio State University, Columbus, Ohio 43210, USA\\}


\begin{abstract}
Do neutrinos have sizable self-interactions? They might. Laboratory constraints are weak, so strong effects are possible in astrophysical environments and the early universe.  Observations with neutrino telescopes can provide an independent probe of neutrino self (``secret") interactions, as the sources are distant and the cosmic neutrino background intervenes.  We define a roadmap for making decisive progress on testing secret neutrino interactions governed by a light mediator.  This progress will be enabled by IceCube-Gen2 observations of high-energy astrophysical neutrinos. Critical to this is our comprehensive treatment of the theory, taking into account previously neglected or overly approximated effects, as well as including realistic detection physics. We show that IceCube-Gen2 can realize the full potential of neutrino astronomy for testing neutrino self-interactions, being sensitive to cosmologically relevant interaction models.  To facilitate forthcoming studies, we release \texttt{nuSIProp}, \mbox{a code that can also be used to study neutrino self-interactions from a variety of sources. \github{ivan-esteban-phys/nuSIprop}}
\end{abstract}

\maketitle

\section{Introduction}
\label{sec:intro}

Neutrinos, ubiquitous but elusive, remain mysterious.  Though many of their properties are known, we still do not know if they might have large self-interactions, also known as neutrino secret interactions ($\nu$SI)~\cite{Choi:1991aa, Acker:1991ej, Acker:1992eh, Beacom:2004yd}. In that scenario, neutrinos interact with a light boson, increasing the neutrino-neutrino scattering rate with respect to that of the standard model.  Laboratory limits, mostly from meson-decay experiments, are relatively weak~\cite{deGouvea:2019qaz}, allowing $\nu$SI to have large effects, potentially explaining short baseline neutrino anomalies~\cite{Dasgupta:2021ies,Asaadi:2017bhx,Chauhan:2018dkd,Smirnov:2021zgn,Dentler:2019dhz,deGouvea:2019qre, Jeong:2018yts} and the muon $g-2$ anomaly~\cite{Bennett:2006fi, Araki:2015mya, Borsanyi:2020mff, Abi:2021gix, Carpio:2021jhu}.

Allowed $\nu$SI would dramatically affect the evolution of systems with high neutrino densities, such as supernovae~\cite{Shalgar:2019rqe} and the early universe~\cite{Bashinsky:2003tk, Beacom:2004yd, Hannestad:2004qu, Hannestad:2005ex, Bell:2005dr}.  While cosmological measurements have robustly established the presence of a radiation background compatible with expectations for the cosmic neutrino background (C$\nu$B)~\cite{Planck:2018vyg, Fields:2019pfx}, its dynamics are poorly constrained.  In fact, Ref.~\cite{Kreisch:2019yzn} found a \emph{preference} for $\nu$SI in the Planck cosmic microwave background (CMB) data that could help alleviate the $H_0$ and $\sigma_8$ tensions~\cite{Planck:2018vyg, Verde:2019ivm} and impact constraints on inflation~\cite{Barenboim:2019tux} (but see Refs.~\cite{Cyr-Racine:2013jua, Archidiacono:2013dua, Lancaster:2017ksf, Oldengott:2017fhy, RoyChoudhury:2020dmd, Huang:2021dba}, where it is argued that polarization data restrict this possibility).  Regardless of the outcome of that debate, it illustrates that a full understanding of possible $\nu$SI is critical to our understanding of the early universe, especially in the upcoming high-precision cosmology era~\cite{EUCLID:2011zbd, Maartens:2015mra, DESI:2016fyo, CMB-S4:2016ple, LSSTDarkEnergyScience:2018jkl}.

\begin{figure*}[t]
    \makebox[\textwidth][c]{
    \includegraphics[width=\textwidth]{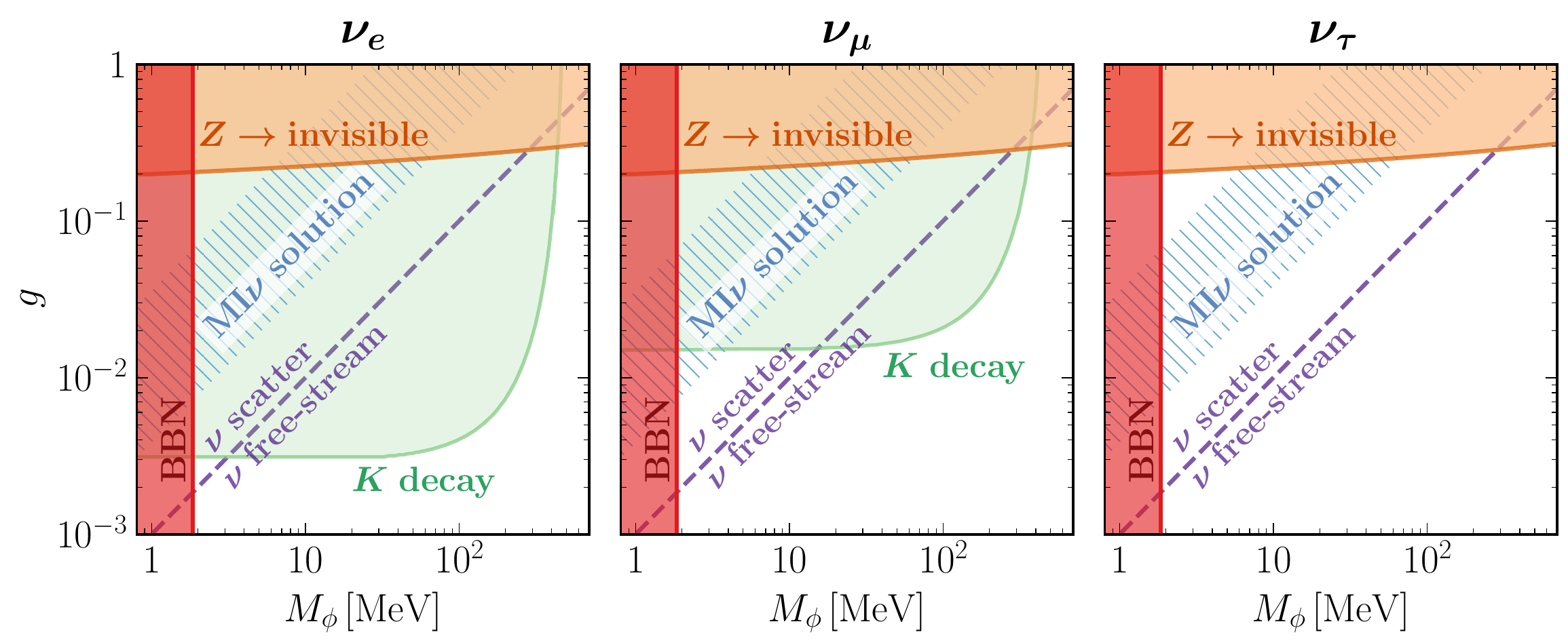}
    }
    \vspace{-0.5cm}
    \caption{Present constraints on neutrino self-interactions, with
      coupling strength $g$ and mediator mass $M_\phi$, for each of the three neutrino flavors (i.e., separately for
      $g_{e \alpha}$ [left], $g_{\mu \alpha}$ [center] and
      $g_{\tau \alpha}$ [right], for $\alpha \in \{e ,\mu, \tau\}$).
      The hatched region is the ``Moderately Interacting
      neutrino'' (MI$\nu$) solution~\cite{Kreisch:2019yzn}, argued to affect CMB observables.
      The dashed purple line is the interaction strength below which cosmic
      neutrinos free-stream as expected at cosmologically relevant times (see text for details). Above this line, our understanding of the early universe would be affected. \emph{As shown, $\nu_\tau$ self-interactions are the least explored, leaving room for
      significant cosmological neutrino effects.}
    } 
    \label{fig:bounds}
    \end{figure*}

\Cref{fig:bounds} shows another essential point about $\nu$SI: while the constraints are strong for $\nu_e$, they are incomplete for $\nu_\mu$ and nearly nonexistent for $\nu_\tau$~\cite{Blinov:2019gcj} (this figure is explained in detail in Sec.~\ref{sec:secret}).  This is why large cosmological effects from $\nu$SI remain allowed.

A new opportunity to probe $\nu$SI has arisen~\cite{Hooper:2007jr, Ng:2014pca, Ioka:2014kca} due to the detection of high-energy neutrinos by IceCube.  The basic idea is that scattering of astrophysical neutrinos with the C$\nu$B en route to Earth redistributes their energies~\cite{Kolb:1987qy}, leading to dips and bumps in the detected spectrum of the diffuse astrophysical neutrino background.  Because we now know that all mass eigenstates have substantial $\nu_\tau$ components~\cite{Esteban:2020cvm, deSalas:2020pgw, Capozzi:2021fjo}, this provides a new tool to explore $\nu_\tau$ self-interactions beyond the reach of laboratory experiments.  However, with the relatively low statistics of IceCube data~\cite{IceCube:2020acn, IceCube:2020wum}, it is hard to realize the full potential of this technique.

In this paper, we define a roadmap for making decisive progress.  The first key to this is the proposed IceCube-Gen2 (Gen2 for brevity) detector~\cite{IceCube-Gen2:2020qha}.  As shown below, its principal advantage relative to IceCube, beyond its better statistics, is its much wider energy range.  The second key is an improved theoretical treatment.  We define the observable signatures of allowed flavor-dependent $\nu$SI, correcting errors and omissions in prior work.  All calculations are done without any approximation, following careful optimizations.  The third key is a realistic treatment of the detector response, using code provided by IceCube.  To make future studies easier, we make our code publicly available \href{https://github.com/ivan-esteban-phys/nuSIprop}{at this URL} \github{ivan-esteban-phys/nuSIprop}.

This paper is organized as follows. In \cref{sec:secret}, we present the theoretical description of $\nu$SI, the relevance of different effects, and the interplay with other experimental observations.  In \cref{sec:data}, we simulate the prospects of Gen2 to probe $\nu$SI. We show that Gen2 can probe the $\nu_\tau$ and other flavor sectors with sensitivity comparable to that of laboratory studies of the $\nu_e$ sector.  Thus, a combination of observables can fully probe the range of $\nu$SI that would significantly affect the dynamics of the cosmic neutrino background.  In \cref{sec:conclusion}, we conclude and highlight future directions.  In 
\cref{sec:appendix}, we give the results of the full $\nu$SI cross-section calculation.

\section{Secret neutrino interactions}
\label{sec:secret}
In this section, we define the specific $\nu$SI models we consider.
\Cref{subsec:model}
introduces the theoretical description and flavor-dependent bounds on
$\nu$SI. \Cref{subsec:dips} specifies the spectral signatures that we
expect given our present understanding of neutrino masses and
mixings. Finally, in \cref{subsec:xsec}, we demonstrate, by computing the full
$\nu$SI cross-section, the relevance of previously ignored effects
both in the theoretical and observed high-energy astrophysical neutrino spectrum.

\subsection{\texorpdfstring{$\nu$}{nu}SI Model and Conceptual Framework}
\label{subsec:model}

We consider $\nu$SI parametrized by the Lagrangian
\begin{align}
\mathcal{L}\supset -\left(\frac{1}{2}\right) \sum_{\alpha,\, \beta} g_{\alpha \beta} \bar{\nu}_\alpha \nu_\beta \phi\,+ \frac{1}{2} \partial_\mu \phi \, \partial^\mu \phi -\frac{1}{2} M_\phi^2 \phi^2 \, ,
\label{eq:Lag}
\end{align}
where $\nu_\alpha$ are neutrino flavor eigenstates ($\alpha \in \{e,\mu, \tau\}$), $g_{\alpha \beta}$ is the interaction strength between flavors $\alpha$ and $\beta$, $\phi$ is the interaction mediator (that for simplicity we assume to be real), $M_\phi$ is its mass, and the $(1/2)$ prefactor is present if neutrinos are Majorana fermions. We write the coupling strength in the flavor basis, as in this basis laboratory constraints are simpler to express.  This is related to the coupling in the mass basis by $g_{ij} = \displaystyle{\sum_{\alpha, \, \beta}g_{\alpha\beta} U_{\alpha i}^* U_{\beta j}}$, where $U$ is the leptonic mixing matrix and $i \in  \{1,2,3\}$.

We assume scalar interactions, as pseudoscalar interactions give the same results at high energies and vector interactions are strongly constrained from laboratory experiments~\cite{Laha:2013xua, Karshenboim:2014tka} and Big Bang Nucleosynthesis~\cite{Huang:2017egl, Blinov:2019gcj}.  Although the Lagrangian in \cref{eq:Lag} is not gauge invariant due to the new couplings affecting neutrinos but not charged leptons, in this paper we are only interested in the low-energy phenomenology effectively described by \cref{eq:Lag}. For possible UV completions that do not generate large couplings to charged leptons, see Refs.~\cite{Blum:2014ewa, Berryman:2018ogk, Blinov:2019gcj, Kelly:2020pcy}. These UV completions might induce additional signatures such as relatively large mixings with sterile neutrinos or a modified Higgs phenomenology, but these are beyond the scope of this paper.  We assume Majorana neutrinos, as for Dirac neutrinos stronger BBN constraints would apply~\cite{Blinov:2019gcj} (see \cref{subsec:dips} for comments on the phenomenology of Dirac neutrinos).

\Cref{fig:bounds} shows that $\nu$SI, particularly in the $\nu_\tau$ sector, are poorly constrained and can affect our understanding of the early universe.  The present constraints (at $2\sigma$) are shown in shaded contours, which come from $Z$-boson decay~\cite{Brdar:2020nbj}, charged kaon decays~\cite{Blinov:2019gcj, Blum:2014ewa, Kelly:2019wow}, and Big Bang Nucleosynthesis (BBN; obtained with the \verb+AlterBBN 2.2+ code~\cite{Arbey:2011nf, Arbey:2018zfh}). To show the scale where early-universe effects become important, the dashed purple line indicates the interaction strength below which cosmological neutrinos free-stream as expected at times relevant to the evolution of the CMB multipoles observed by Planck ($\ell \lesssim 2500$).  We conservatively assume this to be the case if the neutrino self-interaction rate (decreasing as the universe expands) is already below 10\% of the Hubble expansion rate $H$ when the smallest Fourier modes observed by Planck are still well outside the horizon (i.e., for a scale factor ${a = 10 \, k/H}$, with $k$ the comoving wavenumber corresponding to $\ell \sim 2500$).

As an example of $\nu$SI that is presently allowed, the hatched region indicates the Moderately Interacting Neutrino (MI$\nu$) solution~\cite{Kreisch:2019yzn, Blinov:2019gcj}, which has been argued to affect cosmological parameter extraction from CMB data, especially the Hubble constant $H_0$ and the amplitude parameter $\sigma_8$.  Even if the MI$\nu$ solution fades away once more data is accumulated (there seem to be indications in that direction from Planck polarization data~\cite{Cyr-Racine:2013jua, Archidiacono:2013dua, Lancaster:2017ksf, Oldengott:2017fhy, Huang:2021dba}), it will remain important to probe the full parameter space above the purple dashed line.

The weakest $\nu$SI constraints are in the $\nu_\tau$ sector.  Because of flavor mixing, astrophysical neutrinos must always contain a large $\nu_\tau$ component.  In what follows, we explore how $\nu$SI affect astrophysical neutrino propagation, assuming that $\nu$SI apply only in the $\nu_\tau$ sector, i.e., $g_{\alpha \beta} = g \, \delta_{\alpha \tau} \delta_{\beta \tau}$ (our code \texttt{nuSIProp}, though, can handle new couplings in all sectors).  In \cref{sec:data}, we comment how our results in fact provide comparable sensitivity to all three neutrino flavors.

The basic physics of $\nu$SI affecting astrophysical neutrino propagation is as follows~\cite{Kolb:1987qy, Hooper:2007jr, Ng:2014pca, Ioka:2014kca}. En route to Earth, high-energy astrophysical neutrinos may scatter with neutrinos in the C$\nu$B.  (In the standard model, scattering is irrelevant except perhaps at ultra-high energies through the $Z$ resonance~\cite{Weiler1982,Roulet1993}.). As a consequence, high-energy neutrinos are absorbed and lower-energy neutrinos are regenerated.  This leads to unique dips and bumps in the astrophysical neutrino spectrum.

To quantify this description, we follow the comoving differential neutrino number density per unit energy $E_\nu$, $dn/dE_\nu$, as a function of cosmological time $t$.  Since neutrino oscillation lengths
are much smaller than astrophysical scales, flavor oscillations are quickly averaged out; mass eigenstates rapidly decohere too~\cite{Nussinov:1976uw}. We can thus follow the comoving differential density of neutrinos of mass eigenstate $i \in \{1, 2, 3\}$, $dn_i/dE_\nu \equiv \tilde{n}_i(t,E_\nu)$. Its evolution is dictated by a transport equation~\cite{Ng:2014pca, Creque-Sarbinowski:2020qhz}
\begin{align}
\frac{\partial\tilde{n}_i(t,E_\nu)}{\partial t} & =  
\frac{\partial}{\partial{E_\nu}} \left[H(t) \, E_\nu \, \tilde{n}_i(t,E_\nu)\right]+\mathcal{L}_i(t,E_\nu) \notag \label{eq:master} \\
& -
\tilde{n}_i(t,E_\nu) \sum_j n^t_j \sigma_{ij}(E_\nu) \\
& + \sum_{j, \, k, \, l} n^t_j  \int_{E_\nu}^\infty  dE_\nu' \, \tilde{n}_k(t,E_\nu') \frac{d\sigma_{jk \rightarrow il}}{dE_\nu}(E_\nu',E_\nu)\,. \notag
\end{align}
The first term on the right-hand side accounts for the energy
redshift due to cosmological expansion; $H$ is the Hubble
parameter. The second term represents the production rate of high-energy neutrinos
from astrophysical sources: we assume it follows 
the star formation rate~\cite{Hopkins:2006bw, Yuksel:2008cu} as a function of time (a different redshift dependence has little impact on the results as it can not induce dips and bumps in the neutrino spectrum~\cite{DiFranzo:2015qea}), and a power law $\propto E_\nu^{-\gamma}$ with spectral index $\gamma$ as a function of energy.
The first two terms would thus describe the evolution of astrophysical
neutrinos en route to Earth without new physics.  The third term accounts for
absorption due to $\nu$SI, while the fourth term accounts for the subsequent regeneration.  In this equation,
$n^t_i \simeq 2 \times 56\, (1+z)^3\, \text{cm}^{-3} = 8.7 \, (1+z)^3\,  \times
10^{-13} \, \mathrm{eV}^{3}$ is the C$\nu$B density of $\nu_i$, with
$z$ the cosmological redshift, 
$\sigma_{ij}(E_\nu)$ is the absorption cross-section of an incident
$\nu_i$ with energy $E_\nu$ on a target
$\nu_j$, and $\sigma_{jk \rightarrow
  il}(E_\nu', E_\nu)$ is the cross-section for an incident $\nu_j$
with energy $E_\nu'$ on a target $\nu_k$ to generate a
\emph{detectable} $\nu_i$ with energy $E_\nu$, along with a
$\nu_l$. 
For Majorana neutrinos, all final states are detectable; for Dirac neutrinos, neutrinos must be left-handed and antineutrinos right-handed. 

We ignore the neutrino matter potential induced by $\nu$SI~\cite{Lunardini:2000fy,Arguelles:2015dca,Bustamante:2015waa}. \emph{A priori}, it could affect the flavor composition at Earth, but for scalar $\nu$SI it just generates a negligible correction to neutrino masses~\cite{Ge:2018uhz}. For vector $\nu$SI, however, this effect might be measurable at higher energies than the ones we consider if the cosmic neutrino background has a large lepton asymmetry~\cite{Lunardini:2000fy}.

By solving the
differential equation~\eqref{eq:master} with the initial condition
$\tilde{n}_i(z\rightarrow\infty, E_\nu)=0$, we evaluate
${\tilde{n}_i(z=0,E_\nu) \equiv \phi_i(E_\nu)}$, the neutrino flux at
Earth with energy $E_\nu$. The details on how we numerically solve
\cref{eq:master} are given  \href{https://github.com/ivan-esteban-phys/nuSIprop}{at this URL} \github{ivan-esteban-phys/nuSIprop}.
 
\subsection{\texorpdfstring{$\nu$}{nu}SI Effects on the Spectrum}
\label{subsec:dips}

Qualitatively, the effects of $\nu$SI can be understood through the s-channel neutrino-neutrino scattering cross-section (which is dominant, but not to the exclusion of other terms; see \cref{subsec:xsec}),
\begin{align}
\sigma^{s-only}_{ij}= |U_{\tau i}|^2 |U_{\tau j}|^2 \frac{|g|^4}{16\pi}  \frac{s}{(s-M_\phi^2)^2+M_\phi^2 \Gamma^2}\,,
\label{eq:res}
\end{align}
where $s\equiv2 E_\nu m_j$ is the center-of-momentum energy squared, $m_j$ is the mass of $\nu_j$, and $\Gamma = g^2 M_\phi/16\pi$ is the total scalar decay width. (Note that there is a factor of $\sqrt{2}$ difference in our definition of $g$ with respect to Ref.~\cite{Ng:2014pca}.)

At $s=M_\phi^2$, the cross-section is resonantly enhanced, $\sigma_\mathrm{res} \sim |g|^2/M_\phi^2$~\cite{Creque-Sarbinowski:2020qhz}, leading to an enhanced astrophysical neutrino absorption. Since $U_{\tau i} \sim \mathcal{O}(1)$ for all mass eigenstates, the astrophysical spectrum will generically feature multiple absorption dips, located at ${E_\nu \sim M_\phi^2 / (2 m_j)}$ with $j \in \{1,2,3\}$~\cite{Blum:2014ewa, DiFranzo:2015qea, Cherry:2016jol}.

\begin{figure}[t]
 \begin{center}
   \includegraphics[width=\columnwidth]{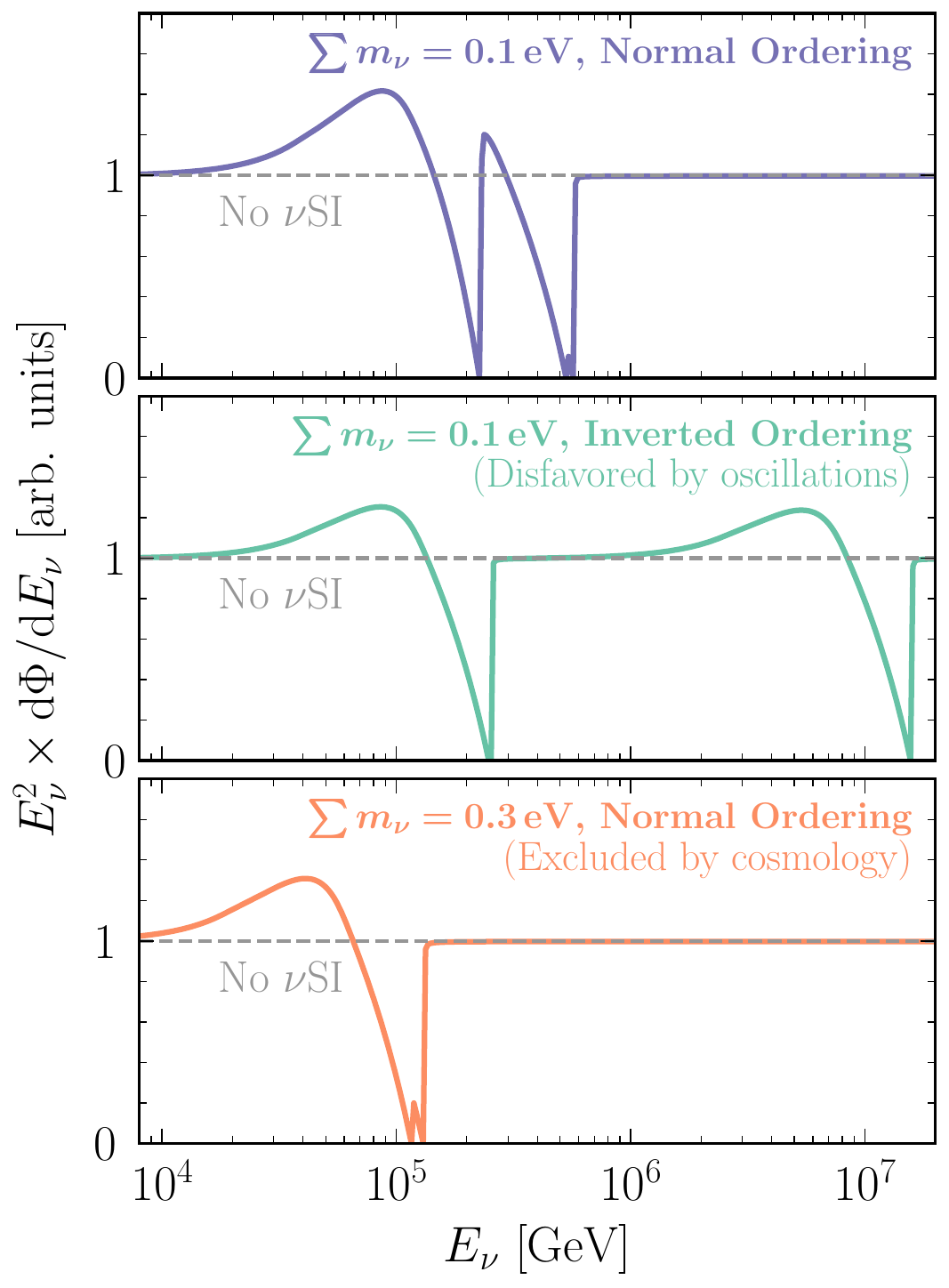}
 \end{center}
 \vspace{-0.5cm}
 \caption{Neutrino flux at Earth assuming  an astrophysical flux $\propto E_\nu^{-2}$ and $\nu_\tau$ self-interactions ($g=0.01$ and $M_\phi = 5 \, \mathrm{MeV}$), for different neutrino mass scenarios. \emph{The scenario in the top panel, favored at $\gtrsim 2.5 \sigma$ over the other two scenarios, would produce two close dips in the astrophysical neutrino spectrum.}}
 \vspace{-0.2cm}
 \label{fig:dips}
\end{figure}   

The separation between these dips, and thus their separate
observability, depends on the neutrino mass spectrum. In the last few
years, there has been a lot of progress in understanding it:
cosmological data bounds the total neutrino mass ${\sum m_\nu < 0.12 \,
  \mathrm{eV}}$ at 95\% CL~\cite{Planck:2018vyg}, and neutrino oscillation data prefer the Normal mass
Ordering (NO), i.e., $m_3 > m_2 > m_1$, over the Inverted mass Ordering
(IO), i.e., $m_2 > m_1 > m_3$, with $\sim 2.5
\sigma$~\cite{Esteban:2020cvm}. (As is standard in neutrino oscillation studies, we denote as $\nu_1$ and $\nu_2$ the eigenstates that are closest in mass, with $m_2 > m_1$. The other eigenstate is denoted as $\nu_3$.) Admittedly, the cosmological bound is
subject to some model dependence
\cite{DiValentino:2019dzu, RoyChoudhury:2019hls, Escudero:2020ped, Esteban:2021ozz}, and the
oscillation preference has lately
weakened~\cite{Esteban:2020cvm}. However, both hints are already
statistically significant and the situation should improve quickly. 
For simplicity, we calculate our results assuming the NO.  As we show below,  the main difference between the NO and the IO appears in present IceCube data, which is not our focus.  For Gen2, we show that it has comparable sensitivity to $\nu$SI for both mass orderings. 

In the NO, $\nu$SI in the $\nu_\tau$ sector
should feature \emph{two} absorption dips, separated in energy by an
$\mathcal{O}(1)$ factor. The reason for this is as follows. First,
$\sum m_\nu < 0.12 \, \mathrm{eV}$, together with the measured neutrino
squared mass splittings $\sqrt{|\Delta m^2_{32}|} \sim \sqrt{|\Delta
  m^2_{31}|} \sim 0.05 \, \mathrm{eV}$, implies that the mass
eigenstates are not degenerate. At the same time, for the NO the spectrum
contains one heavy state and two light states (as $|\Delta m^2_{31}| \gg
\Delta m^2_{21}$, i.e., $m_3 \gg m_1 \sim m_2$); whereas for the IO there
are two heavy states and one light state ($m_2 \sim m_1 \gg
m_3$). Thus, for the same total neutrino mass, the lightest mass for the NO ($m_1$) will typically be much larger than 
the lightest mass for the IO ($m_3$). As a consequence, the NO will generically
imply that light and heavy mass eigenstates are closer to each other, i.e., for the NO the $\nu$SI absorption dips will be closer to each
other. Although in principle there are three absorption dips (one for each mass eigenstate) the smallness of $\Delta m^2_{21}$ implies that two of them are essentially at the same energy unless $0.06 \, \mathrm{eV} \lesssim \sum m_\nu \lesssim 0.07 \, \mathrm{eV}$~\cite{Beacom:2002cb}.

\Cref{fig:dips} shows these features.  We display the all-flavor high-energy astrophysical neutrino flux at Earth obtained by numerically solving \cref{eq:master}.  We plot this as $E_\nu^2 \, \mathrm{d}\phi/\mathrm{d}E_\nu \simeq 2.3^{-1} \, E_\nu \, \mathrm{d}\phi/\mathrm{d}\log_{10} E_\nu$, so that the conserved area under each curve represents the conservation of energy (for the Majorana case, as discussed below). For clarity we have assumed an astrophysical neutrino spectrum $\propto E_\nu^{-2}$ (steeper spectra would make the bumps less visible); below we relax this assumption. We observe the two dips due to neutrino absorption, as well as the two bumps due to the scattering products moving to lower energies. Although absorption mostly takes place at $E_\nu = M_\phi^2 / (2 m_j)$, neutrino energy gets redshifted as the universe expands, extending the dips to lower energies. While the IO also predicts two absorption dips for $\sum m_\nu = 0.1 \, \mathrm{eV}$, they are separated by about two orders of magnitude in neutrino energy. As the present IceCube data covers a rather narrow energy range, this would appear as a \emph{single} dip in the spectrum. Future neutrino telescopes as Gen2 (see \cref{sec:data}) should have a wider energy range and thus $\nu$SI could feature two dips even for the IO.

The fact that we are considering interactions only in the $\nu_\tau$ sector could have an imprint on the flavor composition of the observed high-energy astrophysical neutrino flux.  However, we find the effects to be modest, as shown next.  The Dirac or Majorana nature of neutrinos may also affect the results.  If neutrinos are Dirac fermions, part of the final scattering states will be unobservable (c.f.~\cref{tab:dsigma} in \cref{sec:appendix}). As a consequence, the characteristic $\nu$SI bumps will be smaller. This distinction between Dirac and Majorana neutrinos is expected in any new physics scenario where neutrinos have non-chiral interactions~\cite{Zralek:1997sa}. 

\begin{figure}
\includegraphics[width=\columnwidth]{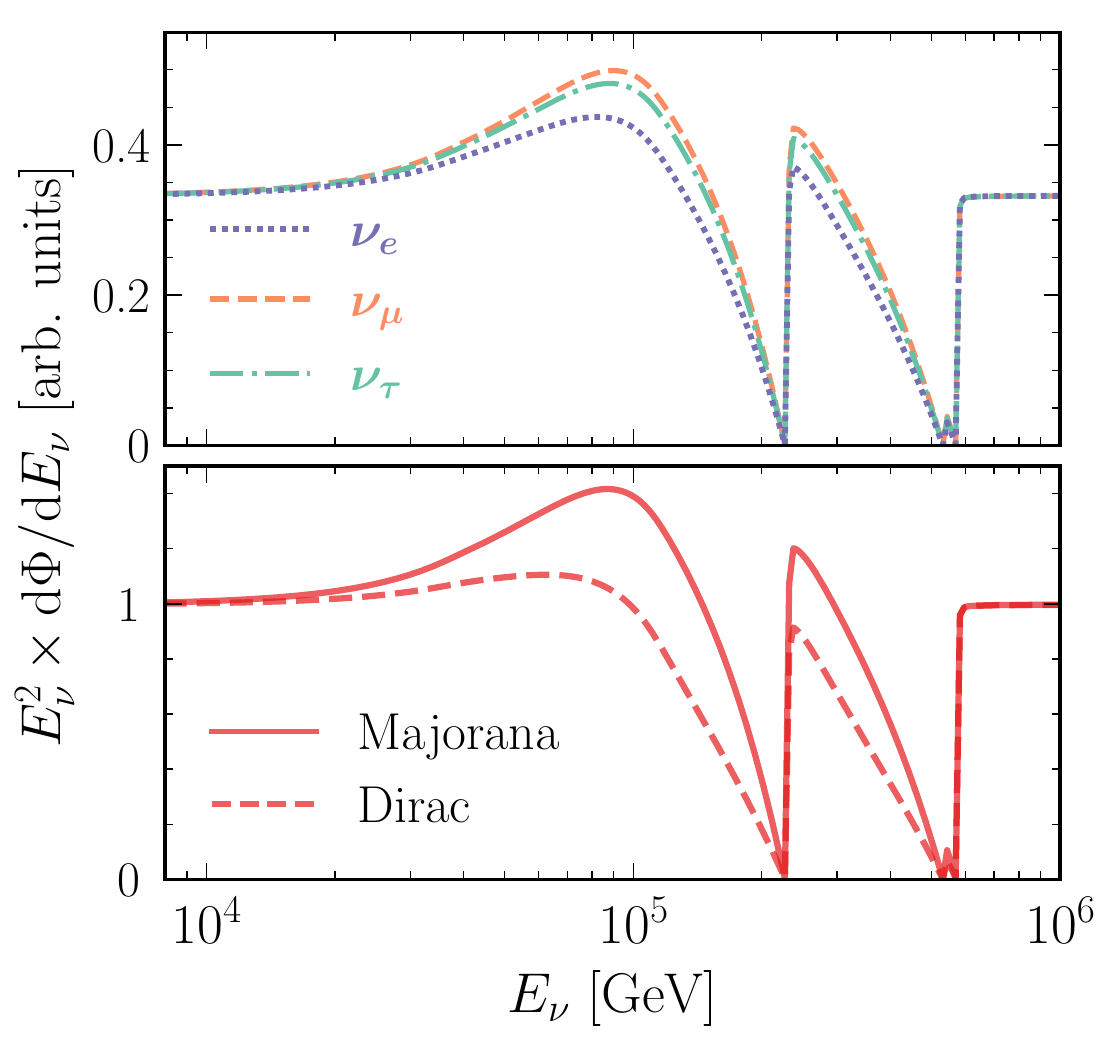}
\vspace{-0.5cm}
\caption{Subleading effects in the neutrino flux at Earth, assuming an astrophysical flux $\propto E_\nu^{-2}$ and $\nu_\tau$ self-interactions ($g=0.01$ and $M_\phi = 5 \, \mathrm{MeV}$), 
  \textbf{Top}: Flux for
  each neutrino flavor, 
  assuming equal production of all mass eigenstates. 
  Within the relevant precision and focusing on the dips,
  \emph{observing the all-flavor flux is enough to explore
  $\nu_\tau$ self-interactions}. 
  \textbf{Bottom}: All-flavor flux,
  contrasting Majorana neutrinos [solid], for  which all scattering products are
  observable; and Dirac neutrinos [dashed], for which some scattering
  products are sterile.
} 
 \vspace{-0.2cm}
 \label{fig:subleading}
\end{figure}   

\Cref{fig:subleading} (top) shows that measuring the all-flavor flux is enough to explore the flavor-dependent $\nu$SI we consider.  We show the high-energy astrophysical neutrino flux at Earth for each neutrino flavor, obtained by numerically solving \cref{eq:master} for $\sum m_\nu = 0.1 \, \mathrm{eV}$ and the NO. Assuming $\nu_\tau$ self-interactions implies that the scattering products will be $\tau$ neutrinos, but since $U_{\tau i} \sim \mathcal{O}(1)$ for all flavors, these will quickly oscillate to an almost flavor-universal composition at Earth. A very good experimental flavor-discrimination, though, might be advantageous in the future as a signature of flavor-dependent $\nu$SI~\cite{Rasmussen:2017ert,Song:2020nfh}.

\Cref{fig:subleading} (bottom) shows the difference between Dirac and Majorana neutrinos. We assume the same physics parameters as in \cref{fig:subleading} (top), but by requiring neutrinos to be Dirac fermions, we observe smaller appearance bumps. This effect can be understood from \cref{eq:Lag,eq:res}: if scattering is mediated through the s-channel resonance, it can be understood as on-shell production of $\phi$ followed by its decay. From \cref{eq:Lag}, we see that the decay products must have opposite chiralities, and so for Dirac neutrinos one of them will be sterile. Since this effect is subleading and, furthermore, constraints on Dirac neutrino self-interactions are rather strong~\cite{Blinov:2019gcj}, we assume Majorana neutrinos for the rest of this paper.

In short, our precise understanding of neutrino properties clearly identifies the experimental signatures we must look for to make the most of the opportunity suggested by \cref{fig:bounds} and explore tau neutrino self-interactions: multiple dips in the astrophysical spectrum with an almost flavor-independent flavor composition.

\subsection{Importance of the Cross-Section Calculation}
\label{subsec:xsec}

\begin{figure}
\centering
 \includegraphics[width=\columnwidth]{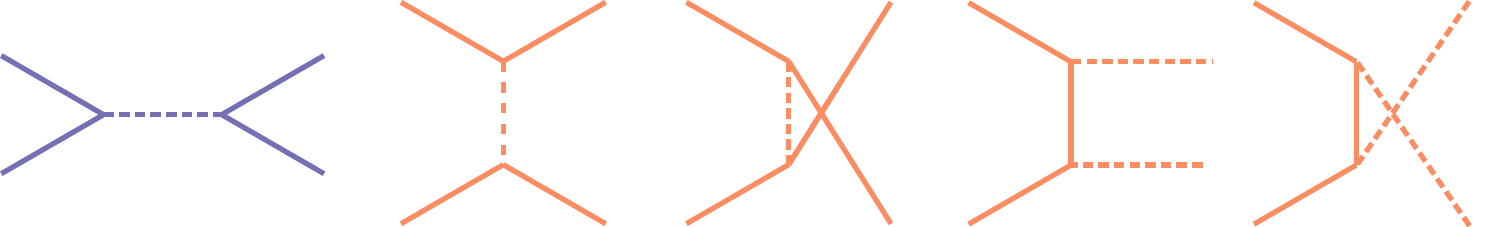}
 \includegraphics[width=\columnwidth]{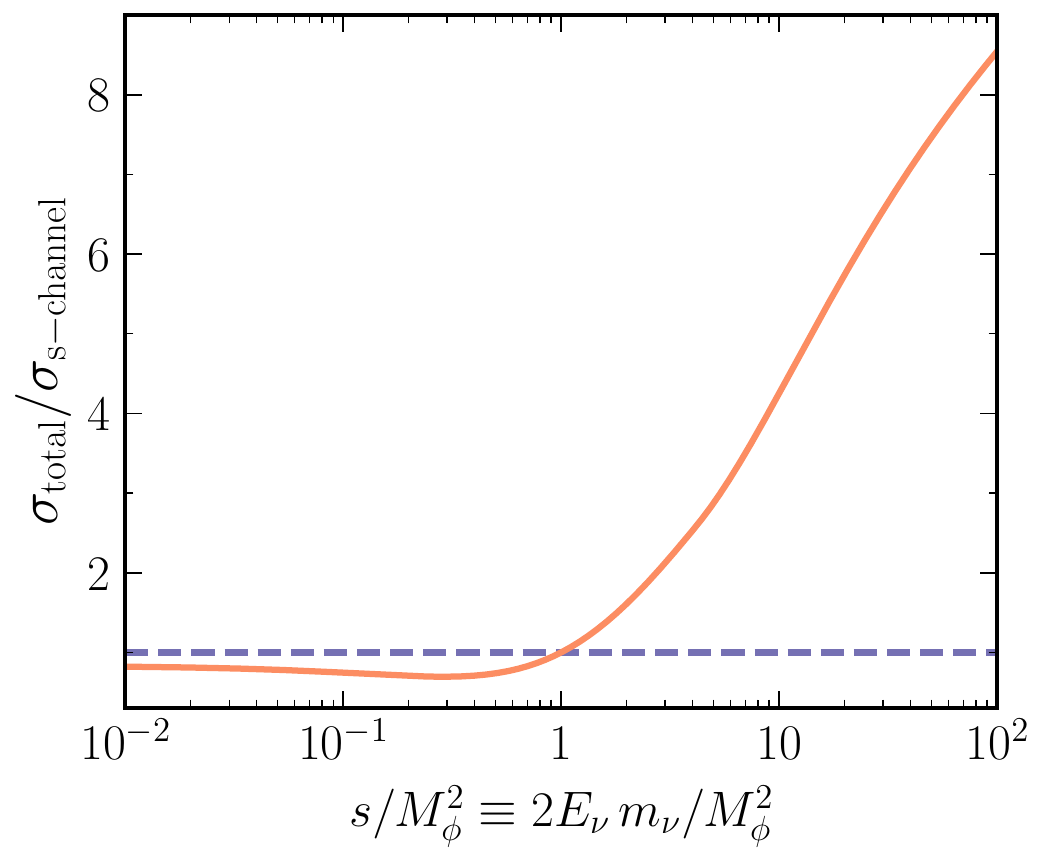}
 \vspace*{-0.75cm}
 \caption{Comparison of the s-channel cross-section with other
   contributions. \textbf{Top}: All diagrams that contribute to neutrino-neutrino scattering at lowest order.
   For double-scalar production, the scalars quickly decay to neutrinos.
   \textbf{Bottom}: Total
   neutrino self-interaction cross-section divided by the s-channel
   contribution as a function of the scaled center-of-momentum energy squared.  \emph{At high energies, the s-channel contribution is subdominant}.}
    \vspace{-0.35cm}
 \label{fig:off-res}
\end{figure}   

\begin{figure}
 \begin{center}
   \includegraphics[width=\columnwidth]{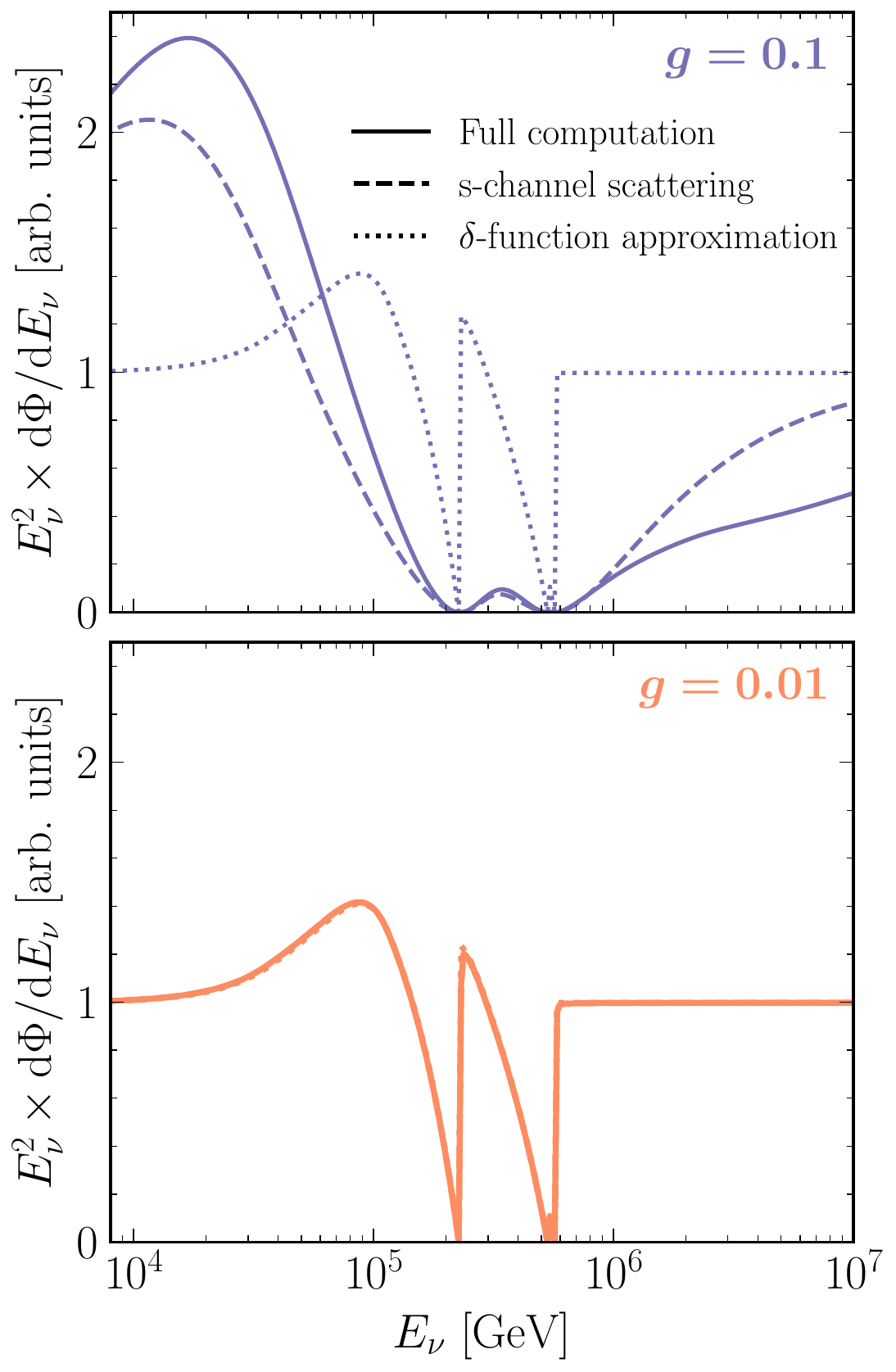}
 \end{center}
 \vspace{-0.6cm}
 \caption{Neutrino flux at Earth in the presence of $\nu$SI,
   using different approximation levels to solve
   \cref{eq:master}. We 
   assume an astrophysical flux $\propto E_\nu^{-2}$ and  $\nu_\tau$ self-interactions ($M_\phi = 5 \, \mathrm{MeV}$).
   \textbf{Top}: For $g=0.1$, a full calculation is needed to obtain correct results.
   \textbf{Bottom}: For $g=0.01$, approximations may be made.
   \emph{To cover the full space of allowed and cosmologically relevant $\nu$SI, a full calculation is needed.}
 }
\vspace*{-0.2cm}
 \label{fig:off-res-flux}
\end{figure}   

In \cref{eq:master}, the neutrino self-interaction cross-section
$\sigma$ parametrizes all the necessary particle physics inputs.  Here we show that approximations made in the literature were unjustified in some important cases.  So far in the literature, only the s-channel contribution (\cref{eq:res}) has been taken into account~\cite{Hooper:2007jr, Ng:2014pca, Bustamante:2020mep, Creque-Sarbinowski:2020qhz, Mazumdar:2020ibx}. Furthermore, as \cref{eq:master} can be computationally expensive to solve, \cref{eq:res} was recently approximated \cite{Creque-Sarbinowski:2020qhz, Mazumdar:2020ibx, Carpio:2021jhu} with a $\delta$-function at $s=M_\phi^2$.

\Cref{fig:off-res} shows that other contributions to the cross section (see \cref{sec:appendix}) are in fact dominant for $s >M_\phi^2$. One can check that for coupling strengths ${g \sim O(0.1)}$ --- according to \cref{fig:bounds}, the couplings that impact our understanding of the early universe --- the astrophysical neutrino interaction probability is large even for $s > M_\phi^2$, and so other scattering channels must be taken into account.

\Cref{fig:off-res-flux} shows that these other scattering channels are quantitatively important.  We show the all-flavor high-energy astrophysical neutrino flux at Earth, obtained by numerically solving
\cref{eq:master} under different approximations for different lines; in all cases we assume $\sum m_\nu = 0.1 \, \mathrm{eV}$ and the NO. As we see, for coupling strengths
$\mathcal{O}(0.1)$, all contributions to the cross-section must be
included to correctly predict the neutrino spectrum, and the
$\delta$-function approximation is unreliable. For $g \sim 0.01$,
on the other hand, all approximations are quantitatively
justified. Note that, for $g = 0.1$, the error introduced by only
including s-channel scattering is smaller than the naive expectation
from \cref{fig:off-res}. This is because, for non
s-channel scattering, the final-state neutrinos tend to have energies
relatively close to the initial neutrino, and so regeneration
partially compensates absorption.

We also show that the $\delta$-function approximation, i.e., only
taking into account the scattering cross-section exactly at the
resonance, gives the same result for both couplings ($g=0.01$ and $g=0.1$) in
\cref{fig:off-res-flux}. This can be understood from the neutrino mean
free path at the resonance, which, ignoring the expansion of the
universe, is~\cite{Creque-Sarbinowski:2020qhz}
\begin{equation}
  \begin{split}
\lambda_\mathrm{res} & \simeq \left(n^t \,
\sigma_\mathrm{res}\right)^{-1} = \left(n^t \, \pi |U_{\tau i}|^2 \frac{g^2}{M_\phi^2}\right)^{-1} \\
& \sim \mathrm{kpc} \times \, g^{-2} \left(\frac{0.1}{|U_{\tau i}|^2}\right) \left(\frac{M_\phi}{5 \, \mathrm{MeV}}\right)^2 \, ,
  \end{split}
  \label{eq:meanfreepath}
\end{equation}
which is much smaller than the typical distances to astrophysical sources, $\sim 1/H_0 \sim \mathrm{Gpc}$, as long as $g \gtrsim 10^{-3}$ (note that oscillation data imply $|U_{\tau i}| \neq 1$~\cite{Esteban:2020cvm, deSalas:2020pgw, Capozzi:2021fjo}). That is --- assuming mediator masses $\mathcal{O}(1$--$10\, \mathrm{MeV})$ --- for all couplings $g \gtrsim 10^{-3}$ all astrophysical neutrinos will be scattered (at $E=E_\mathrm{res}$) regardless of $g$, and so the $\delta$-function approximation will always give the same result. This highlights the importance of carrying out the full theoretical calculation to obtain reliable results.

\begin{figure}[t]
 \begin{center}
   \includegraphics[width=\columnwidth]{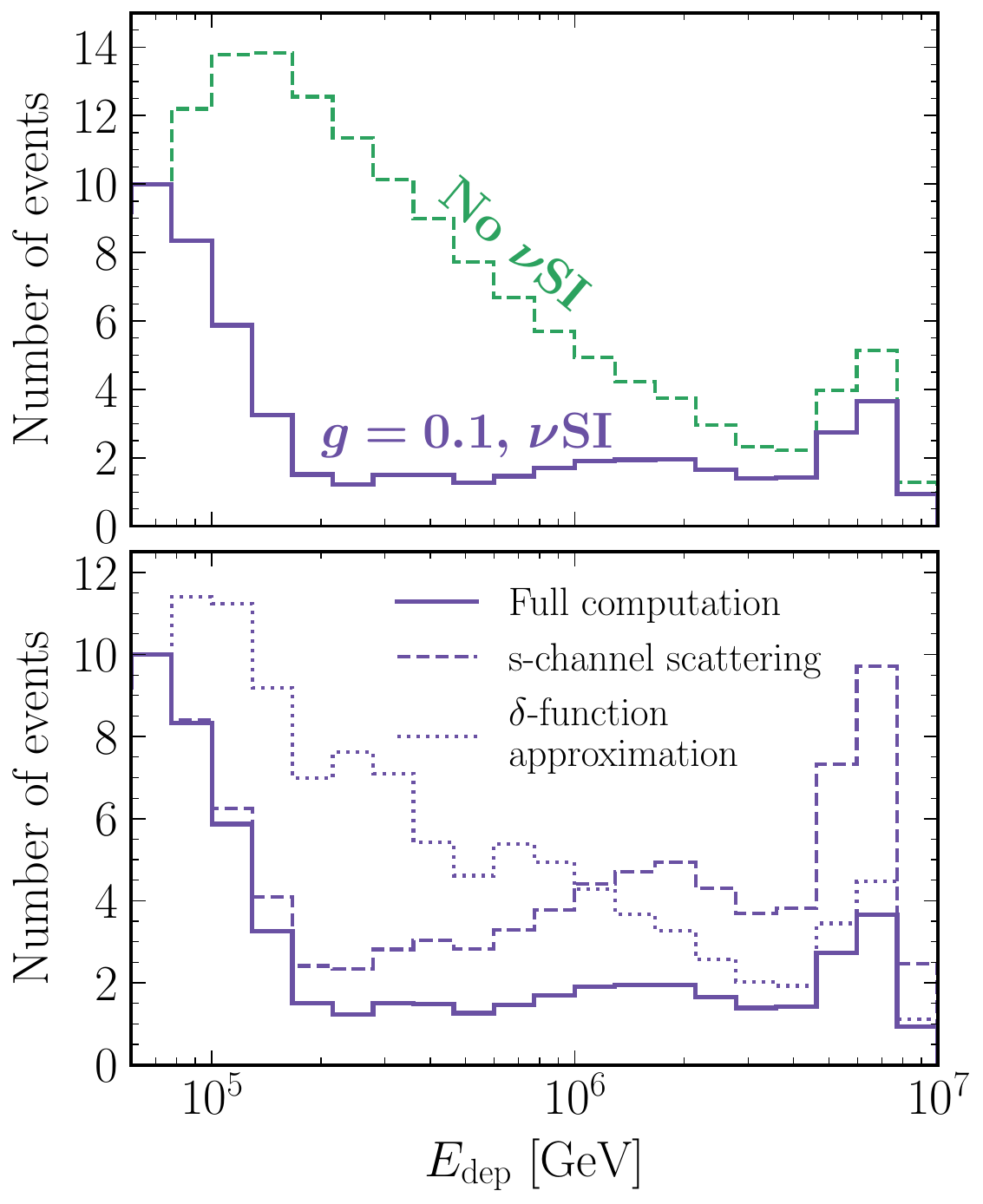}
 \end{center}
 \vspace{-0.6cm}
 \caption{All-flavor astrophysical neutrino events at Earth as a
   function of deposited energy in the detector
   $E_\mathrm{dep}$, assuming a neutrino flux $\propto E_\nu^{-2}$. For the $\nu$SI scenario, we assume $g=0.1$ and $M_\phi = 5 \, \mathrm{MeV}$. \textbf{Top}: Spectrum with and without $\nu$SI. \emph{$\nu$SI leads to quantitatively relevant
   multiple dips and bumps in the detector}. \textbf{Bottom}: $\nu$SI
   spectrum obtained by using different approximations to solve
   \cref{eq:master}. \emph{Unjustified approximations lead to
   incorrect experimental predictions}. 
 } 
 \label{fig:off-res-flux-IC}
 \vspace*{-0.2cm}
\end{figure}   

\Cref{fig:off-res-flux-IC} (top) shows that the effects discussed above are also present after including detection effects. We show in the top panel the expected number of events at IceCube in each bin of deposited energy in the detector, $E_\mathrm{dep}$. (Note that here we plot the number of events per log bin, not the energy-weighted version of that, so now the spectrum is falling.) It is important to use $E_\mathrm{dep}$ instead of $E_\nu$ as the former is different for each flavor, and takes into account the detector energy resolution~\cite{Laha:2013lka}. Using $E_\nu$ would overestimate the detectability of energy-dependent effects.   

To simulate the detector response, we use the official high-energy starting event (HESE) data release \cite{IceCube:2020wum}, which includes all neutrino flavors and interaction topologies and takes into account Earth~\cite{Bustamante:2020mep} and detector effects. We adjust the flux normalization and exposure to predict 10 events in the lowest energy bin as approximately observed by IceCube after 7.5 years (see \cref{sec:data_IC}), and for $\nu$SI we assume $\sum m_\nu = 0.1 \, \mathrm{eV}$ and the NO. We can see the \emph{two} distinct dips, appearing at $E_\mathrm{dep} \sim 2.5 \times 10^5 \, \mathrm{GeV}$ and $E_\mathrm{dep} \sim 5.5 \times 10^5 \, \mathrm{GeV}$, in the presence of $\nu$SI. The effect of the appearance bump is that, for $E_\mathrm{dep} < 2 \times 10^5 \, \mathrm{GeV}$, the neutrino spectrum is steeper than without $\nu$SI. From here, we already foresee a degeneracy between $\nu$SI and the spectral index. Breaking it will require observing the spectrum at a wide neutrino energy range.

The bump at $E_\mathrm{dep} \sim 6 \times 10^6 \, \mathrm{GeV}$ is the Glashow resonance~\cite{Glashow:1960zz}, i.e., resonant production of a $W$ boson in neutrino-electron scattering. This feature has been recently observed by IceCube~\cite{IceCube:2021rpz} and provides a window to the high-energy spectrum.  As shown in \cref{sec:data},  this can help in exploring $\nu$SI.

\Cref{fig:off-res-flux-IC} (bottom) shows that, for couplings $g \sim 0.1$, not including the full cross-section largely overestimates the number of events. We show there the same high-energy astrophysical neutrino flux in the presence of $\nu$SI as in the top panel, but for different levels of approximation when solving \cref{eq:master}. Such approximations were mostly used in prior work to speed up the computations. In our numerical code they are largely unnecessary, as it has been optimized to have a low computational cost. As we make the code publicly available, it can be easily employed in further studies.

\section{IceCube-Gen2: the road to precision neutrino astrophysics}
\label{sec:data}

In this section, we show that Gen2 will be a unprecedented tool for probing $\nu$SI.  In \Cref{sec:data_IC}, we review the present IceCube data and discuss its limitations for exploring $\nu$SI. In \cref{sec:data_gen2}, we discuss how these limitations will be overcome by Gen2. Finally, in \cref{sec:data_analysis}, we carry out a data analysis to quantify the future sensitivity of Gen2 to $\nu$SI as well as the present exclusion by IceCube.

\subsection{Status of IceCube Data}
\label{sec:data_IC}

The IceCube observatory has firmly established the existence of high-energy astrophysical neutrinos by detecting $\mathcal{O}(100)$ neutrinos with $\mathcal{O}(0.1-1)$ PeV  energies. As $\nu$SI introduce an energy-dependent distortion in the astrophysical neutrino spectrum, probing them requires a good reconstruction of the neutrino energy. The HESE dataset~\cite{Abbasi:2020jmh} is particularly appropriate for this, as it only includes neutrinos interacting within the internal part of the detector, thus reducing the amount of energy deposited outside the instrumented volume.

\begin{figure}[t]
 \begin{center}
   \includegraphics[width=\columnwidth]{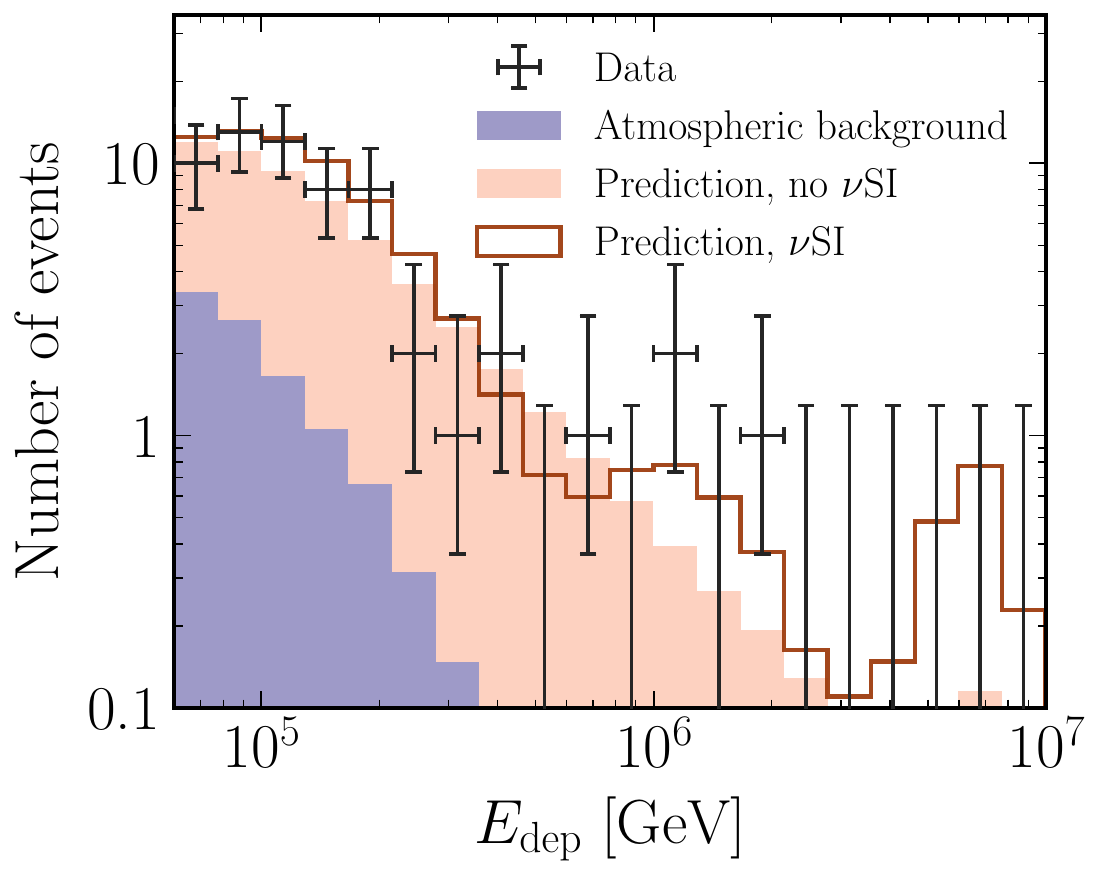}
 \end{center}
 \vspace{-0.5cm}
 \caption{
 HESE 7.5-year data as a function of deposited energy.
 The shaded histograms show the best-fit expectations (with no new physics) for the
 atmospheric background and the astrophysical neutrino signal ($\propto E_\nu^{-2.9}$\cite{Abbasi:2020jmh}). \emph{The data can also be accommodated by $\nu$SI with $g = 0.1$, $M_\phi = 7.5 \, \mathrm{MeV}$, $\sum m_\nu = 0.07 \, \mathrm{eV}$, the NO,
  and a modified astrophysical spectrum ($\propto E_\nu^{-2}$) [solid red line]}.  The large allowed coupling requires taking into account the effects discussed in \cref{subsec:xsec}.
}
 \vspace{-0.2cm}
 \label{fig:spectrumHESE}
\end{figure}

\Cref{fig:spectrumHESE} shows that, intriguingly, the IceCube HESE data suggests some non-trivial spectral features.  High-energy astrophysical acceleration mechanisms generically predict that the neutrino spectrum should follow a power law as a function of energy~\cite{Fermi:1949ee, Gaisser:2016uoy}. This would roughly correspond to a straight line in \cref{fig:spectrumHESE}, but the data shows some deviations. In particular, there seems to be a deficit of events at $E_\mathrm{dep} \sim 5 \times 10^5 \, \mathrm{GeV}$, and an excess at $E_\mathrm{dep} \sim 10^6 \, \mathrm{GeV}$. These features have been interpreted in the literature as hints for $\nu$SI~\cite{Bustamante:2020mep, Mazumdar:2020ibx, Carpio:2021jhu}.

To illustrate this point, we show in  \cref{fig:spectrumHESE} that the spectrum can be fit by nonzero $\nu$SI and a different spectral index.  While the fit is visually better than the no-$\nu$SI case, we caution that the significance is not high (see \cref{sec:data_analysis}).  This $\nu$SI interpretation relies on the presence of \emph{two} separate dips (at $E_\mathrm{dep} \sim 5 \times 10^5 \, \mathrm{GeV}$ and at $E_\mathrm{dep} \sim 3 \times 10^6 \, \mathrm{GeV}$), each with its corresponding bump at lower energies.  We also note that the degeneracy between $\nu$SI parameters and the spectral index could perhaps be exploited to reduce the longstanding tension in the different values of the spectral index deduced from different IceCube datasets~\cite{Aartsen:2015zva, Palladino:2017qda, Abbasi:2020jmh}.  However, this is beyond our scope.  

This discussion identifies two weaknesses of the present IceCube data to explore $\nu$SI: apart from having relatively low statistics, the data only covers about one order of magnitude in neutrino energy. As \cref{fig:dips,fig:off-res,fig:off-res-flux,fig:off-res-flux-IC,fig:subleading} show, $\nu$SI-induced spectral features cover a wide energy range. Furthermore, a wider energy coverage also entails a better understanding of the spectral index that to some extent can be degenerate with $\nu$SI-induced dips, as \cref{fig:spectrumHESE} shows.

\begin{figure}[t]
\centering
\includegraphics[width=\columnwidth]{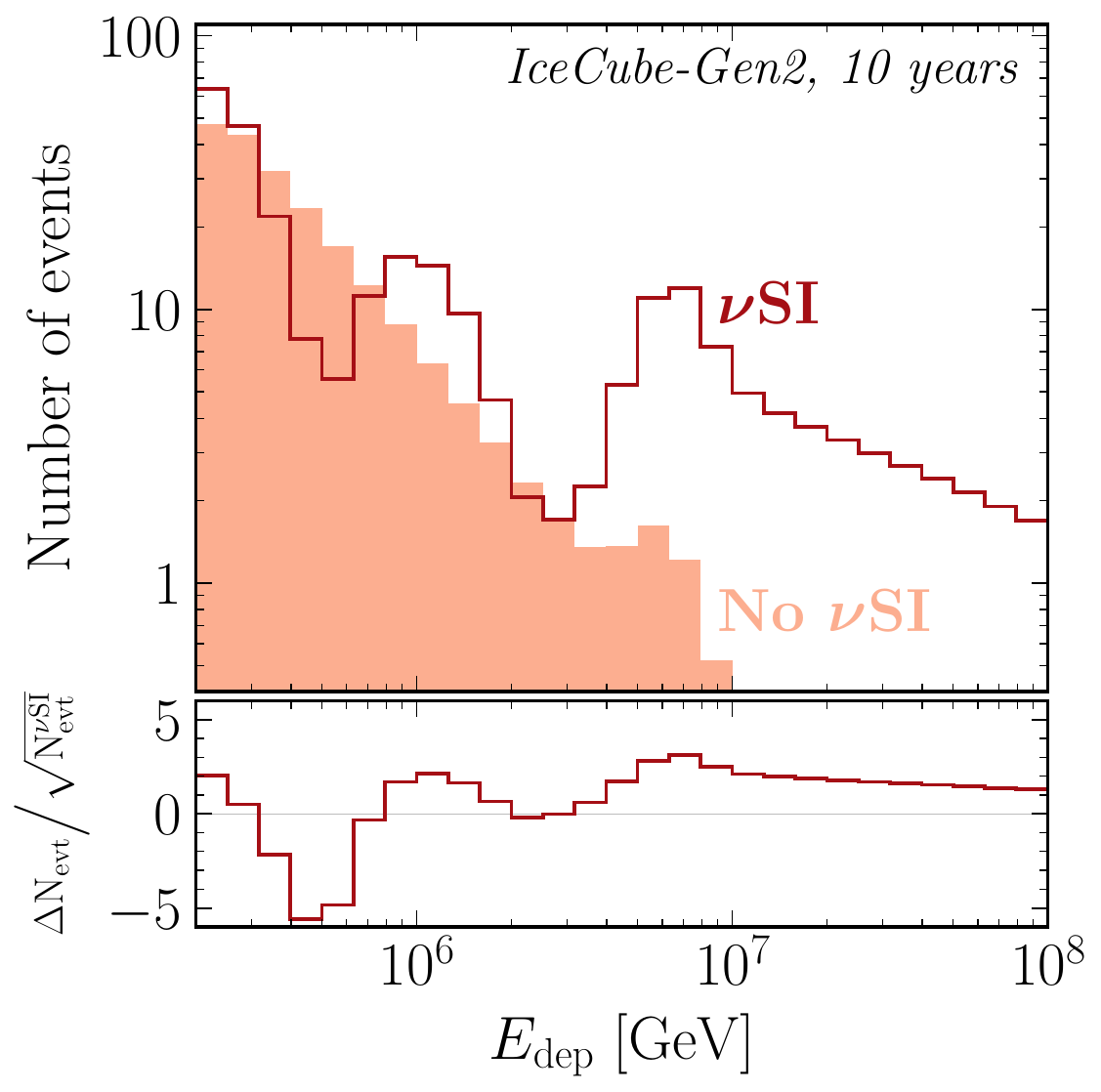}
 \vspace{-0.5cm}
\caption{Spectra for the same parameters as in \cref{fig:spectrumHESE}, but now projecting for Gen2 (note the changes in the axis ranges and the slight change in binning; instead of showing error bars the bottom panel shows the estimated significances), which turns small differences into large ones.  \emph{The increased statistics and the wider energy range reduce the degeneracy with the unknown astrophysical flux} (here $\propto E_\nu^{-2}$ for $\nu$SI and $\propto E_\nu^{-2.9}$ for no $\nu$SI)\emph{, dramatically increasing the sensitivity to $\nu$SI.}} 
 \vspace{-0.2cm}
\label{fig:e_depo_gen2}
\end{figure}

\subsection{Prospects for Gen2}
\label{sec:data_gen2}

Fortunately, Gen2~\cite{IceCube-Gen2:2020qha} will overcome the issues noted above for IceCube.  Its effective volume will be about one order of magnitude larger, so \cref{fig:spectrumHESE} already indicates that Gen2 should observe events at energies up to at least $E_\mathrm{dep} \sim 10^7 \, \mathrm{GeV}$. (Other neutrino telescopes~\cite{KM3Net:2016zxf, Baikal-GVD:2018isr, P-ONE:2020ljt} could have comparable sensitivity as long as their effective volume is as large as that of Gen2 and their energy resolution is good enough.)

To quantify the potential of this future observatory, we carry out a simplified but realistic simulation of Gen2. We take the Gen2 optical effective area from Fig.~25 in Ref.~\cite{IceCube-Gen2:2020qha} that takes into account Earth attenuation, which is significant at Gen2 energies; for simplicity we neglect $\nu_\tau$ regeneration that is less important, specially for steep power laws. We compute the deposited energy distribution following Refs.~\cite{Gandhi:1998ri, Laha:2013lka}, and we assume an energy resolution of 10\% for the deposited energy.  Following Ref.~\cite{IceCube-Gen2:2020qha}, we assume a minimum detectable neutrino energy of 200 TeV, below which the data may contain too many atmospheric background events.

\Cref{fig:e_depo_gen2} shows that Gen2 will indeed be very sensitive to $\nu$SI. We show the expected numbers of events after 10 years of data taking as a function of deposited energy, using the same physics parameters as in \cref{fig:spectrumHESE}. The bottom panel shows the difference between the expected number of events with and without $\nu$SI, divided by the square root of the former, which gives a conservative underestimate of the expected statistical uncertainties. As we see, unlike presently at IceCube, the much wider energy range, together with the increased statistics, will undoubtedly disentangle a power-law spectrum from $\nu$SI with a modified spectral index. To better understand the sensitivity of Gen2, though, we must explore the impact of weaker $\nu$SI.

\begin{figure}[t]
\centering
\includegraphics[width=\columnwidth]{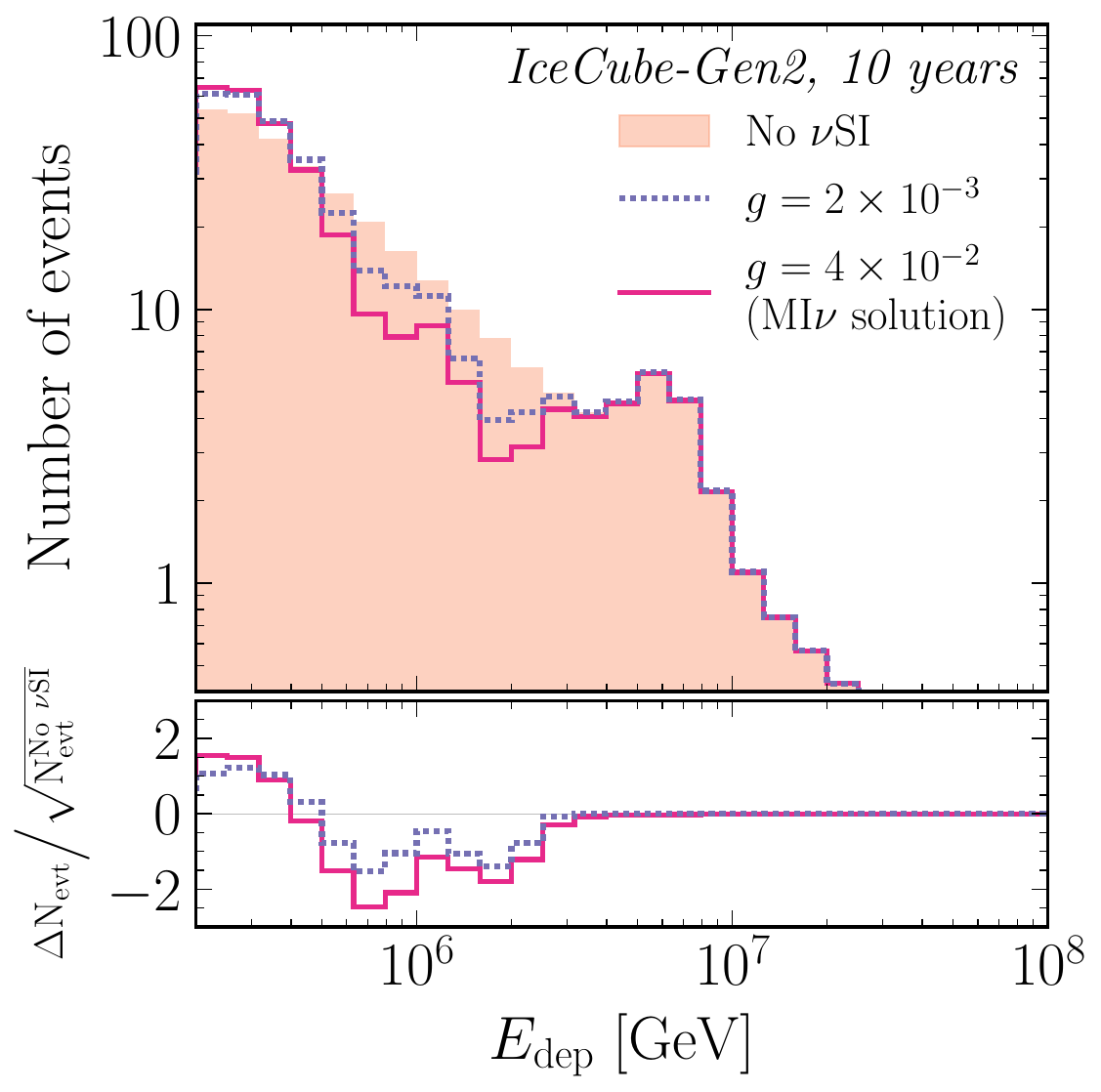}
 \vspace{-0.5cm}
\caption{Expected spectrum of astrophysical neutrino events in Gen2, showing the sensitivity to small couplings.  In all cases, the assumed astrophysical spectrum is $\propto E_\nu^{-2.5}$.  The solid line corresponds to the MI$\nu$ solution for $M_\phi = 10 \, \mathrm{MeV}$, showing that \emph{cosmologically-relevant $\nu$SI would imprint significant features on the Gen2 spectrum}.  The dotted line corresponds to the limiting sensitivity of Gen2, \emph{which cannot be significantly improved upon with neutrino telescopes (see text)}.} 
\label{fig:gen2_10MeV}
\end{figure}

\Cref{fig:gen2_10MeV} quantitatively shows that even relatively weak $\nu$SI would profoundly impact the Gen2 observations. We show the expected spectrum in Gen2, assuming a normalization compatible with current IceCube observations.  In the bottom panel, we show the difference between the number of events with and without $\nu$SI, divided by the expected statistical uncertainties. 
  
The dotted purple line is of particular physical relevance. It corresponds to the coupling at which, for $M_\phi = 10 \, \mathrm{MeV}$, the neutrino mean free path (c.f.~\cref{eq:meanfreepath}) is 1 Gpc, roughly the distance to astrophysical neutrino sources. We expect the effects shown by that line to be close to the weakest detectable $\nu$SI propagation effects: for smaller couplings, the neutrino flux attenuation $\sim e^{-g^2/M_\phi^2 n^t L} \sim 1 - g^2/M_\phi^2 n^t L$, where $L$ is the distance to sources, is hardly different from 1, and uncertainties on the underlying astrophysical flux could start playing an important role. According to the bottom panel, Gen2 should have  enough statistical power to be sensitive to the dotted purple line. It will thus realize the full potential of high-energy astrophysical neutrino propagation for testing $\nu$SI.

The results above also illustrate that the unique spectral features cannot be accommodated by a modified power law and so would be a smoking gun for $\nu$SI. Note that it is not easy for random statistical fluctuations to emulate $\nu$SI, because (1) the separation between the absorption dips is precisely determined by external neutrino oscillation and cosmological data, and (2) $\nu$SI predict a bump with a fixed amplitude just below each energy dip. These features suggest that, should spectral anomalies appear at Gen2, it could be relatively easy to identify them as being due to $\nu$SI or not.

\subsection{\texorpdfstring{$\nu$}{nu}SI sensitivity}
\label{sec:data_analysis}

\begin{figure}[t]
\centering
\includegraphics[width=\columnwidth]{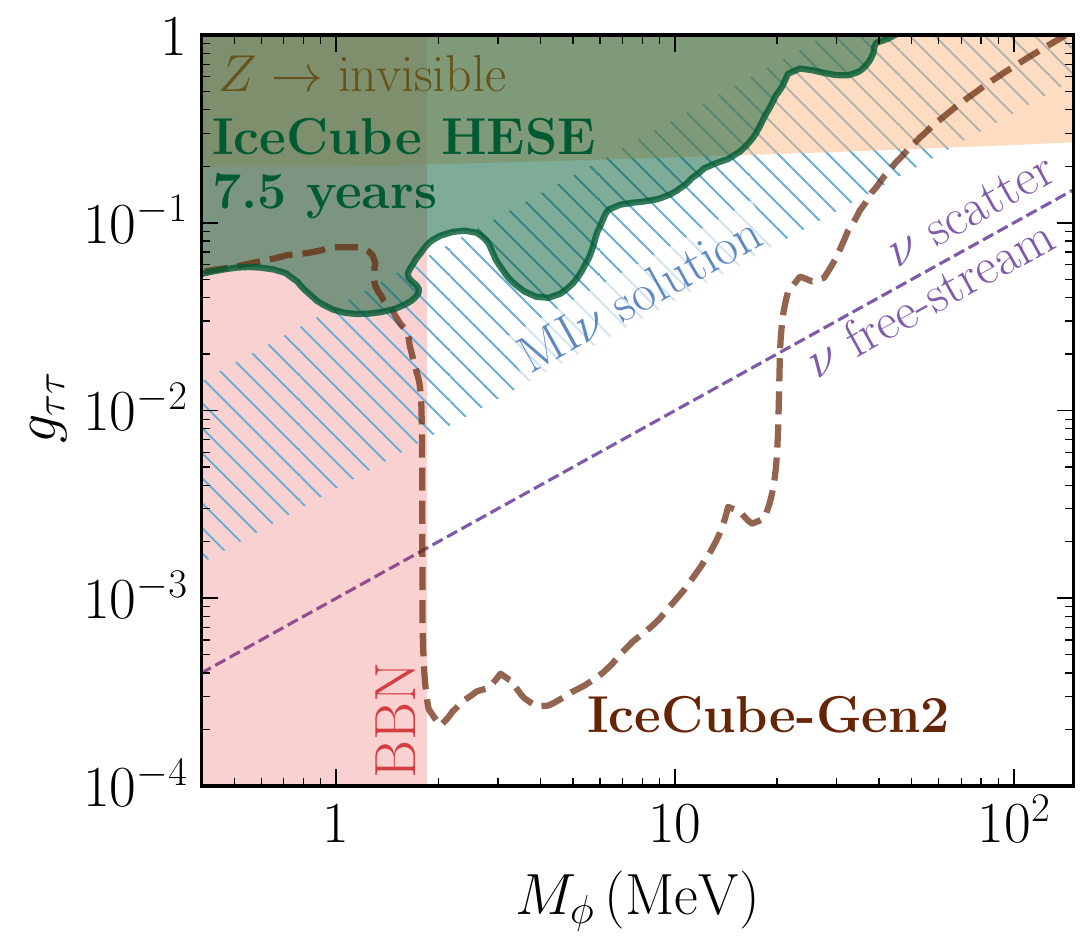}
 \vspace{-0.5cm}
\caption{Present and future sensitivity to $\nu_\tau$
  self-interactions, along with present bounds and cosmologically relevant
  regions (c.f.~\cref{fig:bounds}). The dark green region, including part of the MI$\nu$ region, is excluded
  by IceCube data, and the dashed brown line shows the Gen2 sensitivity (2$\sigma$). 
  \emph{Gen2 will exploit the full potential of testing $\nu$SI with high-energy astrophysical neutrino propagation} (see text)\emph{, being sensitive to
  a large parameter space where neutrinos have a non-trivial cosmological behavior}. The sensitivity to other flavors is comparable.}
\label{fig:contours}
\end{figure}

To quantify the future sensitivity of Gen2, along with the $\nu$SI presently allowed by IceCube, we carry out a data analysis using the theoretical framework discussed in \cref{sec:secret}.

Starting with the present IceCube HESE data, we follow the procedure in the official data release~\cite{Abbasi:2020jmh,thesis}. We assume an initial neutrino flux following a power law with spectral index ${\gamma \in [2, 3]}$ and a normalization ${\phi_{6\nu}(100 \, \mathrm{TeV}) \in [10^{-16}, 10^{-20}] \, \mathrm{GeV}^{-1} \, \mathrm{cm}^{-2} \, \mathrm{s}^{-1} \, \mathrm{sr}^{-1}}$.  In addition to the standard analysis parameters, we include $\nu_\tau$ self-interactions parametrized by the coupling strength $g$, the mediator mass $M_\phi$, and the total neutrino mass $\sum m_\nu$. For $\sum m_\nu$, we include as a prior the cosmological bound coming from CMB and Baryon Acoustic Oscillation data~\cite{Planck:2018vyg}, $\sum m_\nu < 0.12 \, \mathrm{eV}$ (95\% CL). We assume the NO and the neutrino squared mass splittings and mixings from Ref.~\cite{Esteban:2020cvm}.

To forecast the expected sensitivity of Gen2, we generate mock data corresponding to 10 years of exposure at Gen2. As IceCube will keep accumulating data, we also include 15 years of IceCube exposure for deposited energies $E_\mathrm{dep} \in {[6 \times 10^4, \, 2 \times 10^5]\, \mathrm{GeV}}$ that are below the range of Gen2~\cite{IceCube-Gen2:2020qha}. This increases the sensitivity to lower mediator masses. We assume an astrophysical flux following a power law, $\phi_{6 \nu}(E_\nu) = {5 \times 10^{-18} (E_\nu / 100 \, \mathrm{TeV})^{-2.5}\, \mathrm{GeV}^{-1}  \, \mathrm{cm}^{-2} \, \mathrm{s}^{-1} \, \mathrm{sr}^{-1}}$, compatible with present IceCube data. We analyze the mock data assuming $\nu$SI as outlined in the paragraph above (including marginalizing over the spectral index). By the time Gen2 goes online, we expect a precise cosmological determination of $\sum m_\nu$~\cite{EUCLID:2011zbd, Maartens:2015mra, DESI:2016fyo, CMB-S4:2016ple, LSSTDarkEnergyScience:2018jkl}, so we fix this to {0.1 eV}. We use only the HESE spectrum, and do not exploit flavor information. As Ref.~\cite{Song:2020nfh} shows, Gen2 will have  very good flavor discrimination. Including it could enhance the sensitivity to $\nu$SI and probe its flavor structure. 

\Cref{fig:contours} shows that Gen2 will have superb sensitivity, covering a huge range of cosmologically relevant $\nu$SI parameters. The shape of the Gen2 sensitivity curve can be easily
understood. Its best sensitivity to the coupling strength $g$ covers about one order of
magnitude in $M_\phi$ which, for resonant scattering (i.e., $M_\phi^2
\sim 2 E_\nu m_\nu$) corresponds to two orders of magnitude in $E_\nu$. As we see from \cref{fig:e_depo_gen2}, this is
roughly the energy range of Gen2. 

The abrupt sensitivity
decrease for $M_\phi < 2 \, \mathrm{MeV}$ is because, at these
mediator masses and for $\sum m_\nu = 0.1 \, \mathrm{eV}$, the
highest-energy $\nu$SI resonant absorption dip is below the smallest detectable neutrino energy. Gen2 will only be sensitive to such small mediator masses if
neutrino-neutrino scattering outside the resonance is relevant. This,
as \cref{fig:off-res-flux} shows, only happens for coupling strengths
$g \sim 0.1$. These results also illustrate that, since the low-energy statistics is very large (c.f.~\cref{fig:gen2_10MeV}), the presence of a single spectral feature is enough to probe
$\nu$SI.

For mediator masses between $2 \, \mathrm{MeV}$ and $20 \, \mathrm{MeV}$, Gen2 has superb sensitivity. The sensitivity there can be approximated as $ g/M_\phi
\sim$ several times $ 10^{-4} \, \mathrm{MeV}^{-1}$. This corresponds to a
neutrino mean free path at the resonance (c.f.~\cref{eq:meanfreepath}) of
$\lambda_\mathrm{res} \sim \, \mathrm{Gpc} \sim H_0^{-1}$, the
typical distance to astrophysical neutrino sources. For lower couplings,
the neutrino flux attenuation $\sim e^{-g^2/M_\phi^2 n^t L} \sim 1 - g^2/M_\phi^2 n^t L$, where $L$ is the distance to sources, is hardly different from 1. Because Gen2 will have good statistics over a wide energy range, it will be hard to improve upon this sensitivity for mediator masses between $2 \, \mathrm{MeV}$ and $20 \, \mathrm{MeV}$. For example, increasing the statistics by a factor $N$ would only improve the sensitivity by $g \sim N^{1/2}$, even in the ideal case of only having statistical uncertainties.

There are also some local sensitivity improvements for mediator masses of $\sim 4 \,
\mathrm{MeV}$, $\sim 20 \, \mathrm{MeV}$,
and $\sim 30 \, \mathrm{MeV}$. These correspond to the mediator masses at which $\nu$SI spectral features either enter the Gen2 spectrum; or have the same energy as the Glashow resonance, which increases their statistics. 

The sensitivity decreases again for $M_\phi \gtrsim 20 \,
\mathrm{MeV}$, the mediator masses above which the highest-energy $\nu$SI
dip is at energies $E_\mathrm{dep} \gtrsim 10^7 \,
\mathrm{GeV}$. Although there are still other $\nu$SI features in the
spectrum, the poor statistics at high energies dramatically reduces the
sensitivity. For these mediator masses, in contrast to $2 \, \mathrm{MeV} \lesssim M_\phi \lesssim 20 \, \mathrm{MeV}$, additional statistics from higher energy detectors could significantly increase the sensitivity to $\nu$SI. Note that this particular mediator mass above which Gen2
loses sensitivity depends on the spectral index, which sets the
highest energy neutrinos that can be detected.

We have systematically explored how our results depend on the input choices. In short, changing the inputs within reasonable ranges leads to results that are similar enough; hence, \cref{fig:contours} is sufficient to represent the Gen2 sensitivity. Assuming the IO instead of the NO would keep one of the spectral features unaffected (c.f. \cref{fig:dips}), which, as discussed above, is enough to probe $\nu$SI. The assumed total neutrino mass or squared mass splittings do not change the results either, as present oscillation and cosmological data are precise enough to fix the location of the relevant $\nu$SI spectral features~\cite{Beacom:2002cb}. Finally, the assumed spectral index sets the largest energy detectable by Gen2 and thus the largest value of $M_\phi$ that can be explored, but we have checked that changing it to 2.0 or 3.0 only changes the higher end of the Gen2 sensitivity by about a factor of 2 in $M_\phi$. As we discuss below, the higher end of the Gen2 sensitivity can be complemented with higher-energy observatories.

\cref{fig:contours} also shows the present IceCube HESE exclusion region. It is more powerful than BBN and $Z$ decay bounds in some
regions of the $\nu$SI parameter
space, and complementary in others. The explored coupling range
already indicates the relevance of the theoretical effects we
discuss in \cref{sec:secret}: we have checked that, if we ignored them, the IceCube excluded region
would be quantitatively different.

Finally, we note that, unlike in prior work~\cite{Bustamante:2020mep,Mazumdar:2020ibx,Carpio:2021jhu}, we do not find any statistically
significant indication in favor of $\nu$SI in present IceCube data. In particular, despite being described by more parameters, $\nu$SI are
only preferred over a simple power law neutrino spectrum at the $\sim
1 \sigma$ level.
This difference may stem from our more precise theoretical treatment described in
\cref{sec:secret}, our use of a more recent data set, our precise treatment of detector effects that matches that used by IceCube, or a combination of these effects.

Both \cref{fig:contours,fig:gen2_10MeV} illustrate that
Gen2 will be a very powerful tool to explore $\nu$SI. We expect it to
be sensitive to the full range of the presently allowed MI$\nu$
solution and to almost the full parameter space for which
cosmological neutrinos do not free-stream. The remaining allowed region could be explored with higher-energy observatories including Gen2-radio~\cite{GRAND:2015uko, ARA:2019wcf, IceCube-Gen2:2020qha, Abarr:2020bjd, POEMMA:2020ykm,  Prohira:2021vvn, Fiorillo:2020jvy, Fiorillo:2020zzj}. Furthermore,
for mediator masses between 2 and 20 MeV, Gen2 would realize the full potential of neutrino astronomy
to test $\nu$SI, even competing with 
the strongest laboratory probes (see meson-decay limits in \cref{fig:bounds}) and setting the strongest bound on $\nu$SI. Since the presence of a single spectral feature is enough to probe
$\nu$SI at Gen2, we expect the constraints from \cref{fig:contours} to apply with comparable strength to $\nu_e$ and $\nu_\mu$. This future observatory will be a unique
bridge between the cosmological and laboratory quests to understand
if neutrinos have large self-interactions.

\section{Conclusions and Outlook}
\label{sec:conclusion}
We return to our opening question: Do neutrinos have sizable self-interactions?  In the laboratory, this is impossible to answer  through scattering and is not adequately constrained through particle decays.  But this question is of central importance to particle theory, as neutrinos allow unique tests of new physics.  And it is of central importance to cosmology, as allowed $\nu$SI parameters would indicate the presence of strong self-scattering in the early universe. New techniques to probe $\nu$SI are needed, especially in the $\nu_\tau$ sector.  These may lead to discoveries that should be tested in multiple ways, or to limits that will improve our abilities to search for physics beyond the standard models of particle physics and cosmology.

A way forward is to look for signatures of scattering with neutrinos in the C$\nu$B, which leads to characteristic features in the spectrum of astrophysical neutrinos at Earth~\cite{Kolb:1987qy, Hooper:2007jr, Ng:2014pca, Ioka:2014kca}.  This has become newly promising now that IceCube has detected $\mathcal{O}(100)$ events at energies larger than 60 TeV~\cite{IceCube:2020acn, IceCube:2020wum}.

In this paper, we take advantage of the proposed Gen2 detector, develop a comprehensive theoretical treatment, and make predictions that include realistic experimental effects.  We benefit from the improved knowledge of the neutrino mass spectrum: the measurements of the total neutrino mass in cosmology and of the neutrino mass ordering in the laboratory shape the experimental signatures that should be looked for. We also demonstrate that common approximations in the literature were unjustified in some important cases.

Our primary result is in \cref{fig:contours}: Gen2 will significantly improve the sensitivity to $\nu$SI, realizing the full potential of high-energy neutrino astronomy for testing $\nu$SI in propagation. It can probe a huge range of parameters where neutrinos do not free-stream as expected in the early universe. This includes the full parameter space relevant for the Moderately Interacting neutrino solution~\cite{Kreisch:2019yzn,Blinov:2019gcj}. And, as discussed above, modifying the analysis inputs only induces slight sensitivity changes. At its best sensitivity, Gen2 will overcome laboratory constraints and become the strongest probe of neutrino self-interactions \emph{between any flavors}.   The future quest for $\nu$SI discovery will not remain bounded to $\tau$ neutrinos as it is today.

We make our code publicly available so that our formalism can be straightforwardly applied to explore different parts of the parameter space with various sources. Some examples include the diffuse supernova neutrino background or cosmogenic neutrinos. 

Should a signal appear at Gen2, there will be plenty of opportunities to test it. 
A key observable is the flavor composition (see, e.g., Ref.~\cite{Song:2020nfh}).  Another is exploiting point sources.  The main purpose of Gen2 is to resolve individual neutrino  sources~\cite{IceCube-Gen2:2020qha}; any detection would be highly valuable to explore $\nu$SI~\cite{Kelly:2018tyg}. The reason is that nearby sources should not be affected by $\nu$SI and could provide a better understanding of the high-energy astrophysical neutrino spectrum. The appearance of spectral signatures in far but not near sources would be a smoking gun for $\nu$SI. In addition, the scattering of neutrinos en route to the Earth could introduce measurable time delays~\cite{Ng:2014pca}.  In addition to the Gen2 optical array that we considered, the Gen2 radio array and other neutrino observatories can greatly extend the reach of our work by detecting higher-energy neutrinos~\cite{GRAND:2015uko, ARA:2019wcf, IceCube-Gen2:2020qha, Abarr:2020bjd, POEMMA:2020ykm,  Prohira:2021vvn, Fiorillo:2020jvy, Fiorillo:2020zzj}.  Finally, hints for $\nu$SI at Gen2 could leave signatures in astrophysical and cosmological observables.

As neutrino physics enters the precision era, the properties of these ghostly particles will be scrutinized better than ever. In this paper, we provide the necessary particle-physics framework as well as the phenomenological roadmap to make the most out of neutrino self-interaction measurements in present and next-generation neutrino telescopes. Improvements in  understanding high-energy astrophysical sources and further experimental sensitivity studies will  enhance this progress. This will open a window into understanding whether neutrinos have sizable self-interactions, providing insight about physics beyond the standard model and the evolution of the early universe.

\tocless{\section*{Acknowledgements}}

We are grateful for helpful discussions with Dan Hooper, Marc Kamionkowski, Siddhartha Karmakar, Gordan Krnjaic, Jordi Salvado, Ian Shoemaker, Todd Thompson, Xun-Jie Xu, and especially Markus Ahlers, Carlos Argüelles, Mauricio Bustamante, Pilar Coloma, M.C. Gonzalez-Garcia, Shirley Li, Kevin Kelly, Austin Schneider, and Irene Tamborra.  J.F.B. is supported by National Science Foundation grant No.\ PHY-2012955.  Fermilab is operated by the Fermi  Research  Alliance,  LLC  under  contract  No.\ DE-AC02-07CH11359 with the United States Department of Energy. 

\bibliographystyle{JHEP}
\tocless\bibliography{refs}

\providecommand{\href}[2]{#2}\begingroup\raggedright\begin{thebibliography}{100}

\bibitem{Choi:1991aa}
K.~Choi and A.~Santamaria, {\it {17-KeV neutrino in a singlet - triplet majoron
  model}},  {\em Phys. Lett. B} {\bf 267} (1991) 504--508.

\bibitem{Acker:1991ej}
A.~Acker, S.~Pakvasa, and J.~T. Pantaleone, {\it {Decaying Dirac neutrinos}},
  {\em Phys. Rev. D} {\bf 45} (1992) 1--4.

\bibitem{Acker:1992eh}
A.~Acker, A.~Joshipura, and S.~Pakvasa, {\it {A Neutrino decay model, solar
  anti-neutrinos and atmospheric neutrinos}},  {\em Phys. Lett. B} {\bf 285}
  (1992) 371--375.

\bibitem{Beacom:2004yd}
J.~F. Beacom, N.~F. Bell, and S.~Dodelson, {\it {Neutrinoless universe}},  {\em
  Phys. Rev. Lett.} {\bf 93} (2004) 121302,
  [\href{http://www.arxiv.org/abs/astro-ph/0404585}{{\tt astro-ph/0404585}}].

\bibitem{deGouvea:2019qaz}
A.~de~Gouv\^ea, P.~S.~B. Dev, B.~Dutta, T.~Ghosh, T.~Han, and Y.~Zhang, {\it
  {Leptonic Scalars at the LHC}},  {\em JHEP} {\bf 07} (2020) 142,
  [\href{http://www.arxiv.org/abs/1910.01132}{{\tt 1910.01132}}].

\bibitem{Dasgupta:2021ies}
B.~Dasgupta and J.~Kopp, {\it {Sterile Neutrinos}},
  \href{http://www.arxiv.org/abs/2106.05913}{{\tt 2106.05913}}.

\bibitem{Asaadi:2017bhx}
J.~Asaadi, E.~Church, R.~Guenette, B.~J.~P. Jones, and A.~M. Szelc, {\it {New
  light Higgs boson and short-baseline neutrino anomalies}},  {\em Phys. Rev.
  D} {\bf 97} (2018), no.~7 075021,
  [\href{http://www.arxiv.org/abs/1712.08019}{{\tt 1712.08019}}].

\bibitem{Chauhan:2018dkd}
B.~Chauhan and S.~Mohanty, {\it {Signature of light sterile neutrinos at
  IceCube}},  {\em Phys. Rev. D} {\bf 98} (2018), no.~8 083021,
  [\href{http://www.arxiv.org/abs/1808.04774}{{\tt 1808.04774}}].

\bibitem{Smirnov:2021zgn}
A.~Y. Smirnov and V.~B. Valera, {\it {Resonance refraction and neutrino
  oscillations}},  \href{http://www.arxiv.org/abs/2106.13829}{{\tt
  2106.13829}}.

\bibitem{Dentler:2019dhz}
M.~Dentler, I.~Esteban, J.~Kopp, and P.~Machado, {\it {Decaying Sterile
  Neutrinos and the Short Baseline Oscillation Anomalies}},  {\em Phys. Rev. D}
  {\bf 101} (2020), no.~11 115013,
  [\href{http://www.arxiv.org/abs/1911.01427}{{\tt 1911.01427}}].

\bibitem{deGouvea:2019qre}
A.~de~Gouv\^ea, O.~L.~G. Peres, S.~Prakash, and G.~V. Stenico, {\it {On The
  Decaying-Sterile Neutrino Solution to the Electron (Anti)Neutrino Appearance
  Anomalies}},  {\em JHEP} {\bf 07} (2020) 141,
  [\href{http://www.arxiv.org/abs/1911.01447}{{\tt 1911.01447}}].

\bibitem{Jeong:2018yts}
Y.~S. Jeong, S.~Palomares-Ruiz, M.~H. Reno, and I.~Sarcevic, {\it {Probing
  secret interactions of eV-scale sterile neutrinos with the diffuse supernova
  neutrino background}},  {\em JCAP} {\bf 06} (2018) 019,
  [\href{http://www.arxiv.org/abs/1803.04541}{{\tt 1803.04541}}].

\bibitem{Bennett:2006fi}
{\bf Muon g-2} {\bf Collaboration}, G.~W. Bennett {\em et~al.}, {\it {Final
  Report of the Muon E821 Anomalous Magnetic Moment Measurement at BNL}},  {\em
  Phys. Rev. D} {\bf 73} (2006) 072003,
  [\href{http://www.arxiv.org/abs/hep-ex/0602035}{{\tt hep-ex/0602035}}].

\bibitem{Araki:2015mya}
T.~Araki, F.~Kaneko, T.~Ota, J.~Sato, and T.~Shimomura, {\it {MeV scale
  leptonic force for cosmic neutrino spectrum and muon anomalous magnetic
  moment}},  {\em Phys. Rev. D} {\bf 93} (2016), no.~1 013014,
  [\href{http://www.arxiv.org/abs/1508.07471}{{\tt 1508.07471}}].

\bibitem{Borsanyi:2020mff}
S.~Borsanyi {\em et~al.}, {\it {Leading hadronic contribution to the muon
  magnetic moment from lattice QCD}},  {\em Nature} {\bf 593} (2021), no.~7857
  51--55, [\href{http://www.arxiv.org/abs/2002.12347}{{\tt 2002.12347}}].

\bibitem{Abi:2021gix}
B.~Abi {\em et~al.}, {\it {Measurement of the Positive Muon Anomalous Magnetic
  Moment to 0.46 ppm}},  \href{http://www.arxiv.org/abs/2104.03281}{{\tt
  2104.03281}}.

\bibitem{Carpio:2021jhu}
J.~A. Carpio, K.~Murase, I.~M. Shoemaker, and Z.~Tabrizi, {\it {High-energy
  cosmic neutrinos as a probe of the vector mediator scenario in light of the
  muon $g-2$ anomaly and Hubble tension}},
  \href{http://www.arxiv.org/abs/2104.15136}{{\tt 2104.15136}}.

\bibitem{Shalgar:2019rqe}
S.~Shalgar, I.~Tamborra, and M.~Bustamante, {\it {Core-collapse supernovae
  stymie secret neutrino interactions}},  {\em Phys. Rev. D} {\bf 103} (2021),
  no.~12 123008, [\href{http://www.arxiv.org/abs/1912.09115}{{\tt
  1912.09115}}].

\bibitem{Bashinsky:2003tk}
S.~Bashinsky and U.~Seljak, {\it {Neutrino perturbations in CMB anisotropy and
  matter clustering}},  {\em Phys. Rev. D} {\bf 69} (2004) 083002,
  [\href{http://www.arxiv.org/abs/astro-ph/0310198}{{\tt astro-ph/0310198}}].

\bibitem{Hannestad:2004qu}
S.~Hannestad, {\it {Structure formation with strongly interacting neutrinos -
  Implications for the cosmological neutrino mass bound}},  {\em JCAP} {\bf 02}
  (2005) 011, [\href{http://www.arxiv.org/abs/astro-ph/0411475}{{\tt
  astro-ph/0411475}}].

\bibitem{Hannestad:2005ex}
S.~Hannestad and G.~Raffelt, {\it {Constraining invisible neutrino decays with
  the cosmic microwave background}},  {\em Phys. Rev. D} {\bf 72} (2005)
  103514, [\href{http://www.arxiv.org/abs/hep-ph/0509278}{{\tt
  hep-ph/0509278}}].

\bibitem{Bell:2005dr}
N.~F. Bell, E.~Pierpaoli, and K.~Sigurdson, {\it {Cosmological signatures of
  interacting neutrinos}},  {\em Phys. Rev. D} {\bf 73} (2006) 063523,
  [\href{http://www.arxiv.org/abs/astro-ph/0511410}{{\tt astro-ph/0511410}}].

\bibitem{Planck:2018vyg}
{\bf Planck} {\bf Collaboration}, N.~Aghanim {\em et~al.}, {\it {Planck 2018
  results. VI. Cosmological parameters}},  {\em Astron. Astrophys.} {\bf 641}
  (2020) A6, [\href{http://www.arxiv.org/abs/1807.06209}{{\tt 1807.06209}}].

\bibitem{Fields:2019pfx}
B.~D. Fields, K.~A. Olive, T.-H. Yeh, and C.~Young, {\it {Big-Bang
  Nucleosynthesis after Planck}},  {\em JCAP} {\bf 03} (2020) 010,
  [\href{http://www.arxiv.org/abs/1912.01132}{{\tt 1912.01132}}]. [Erratum:
  JCAP 11, E02 (2020)].

\bibitem{Kreisch:2019yzn}
C.~D. Kreisch, F.-Y. Cyr-Racine, and O.~Dor\'e, {\it {Neutrino puzzle:
  Anomalies, interactions, and cosmological tensions}},  {\em Phys. Rev. D}
  {\bf 101} (2020), no.~12 123505,
  [\href{http://www.arxiv.org/abs/1902.00534}{{\tt 1902.00534}}].

\bibitem{Verde:2019ivm}
L.~Verde, T.~Treu, and A.~G. Riess, {\it {Tensions between the Early and the
  Late Universe}},  {\em Nature Astron.} {\bf 3} (7, 2019) 891,
  [\href{http://www.arxiv.org/abs/1907.10625}{{\tt 1907.10625}}].

\bibitem{Barenboim:2019tux}
G.~Barenboim, P.~B. Denton, and I.~M. Oldengott, {\it {Constraints on inflation
  with an extended neutrino sector}},  {\em Phys. Rev. D} {\bf 99} (2019),
  no.~8 083515, [\href{http://www.arxiv.org/abs/1903.02036}{{\tt 1903.02036}}].

\bibitem{Cyr-Racine:2013jua}
F.-Y. Cyr-Racine and K.~Sigurdson, {\it {Limits on Neutrino-Neutrino Scattering
  in the Early Universe}},  {\em Phys. Rev. D} {\bf 90} (2014), no.~12 123533,
  [\href{http://www.arxiv.org/abs/1306.1536}{{\tt 1306.1536}}].

\bibitem{Archidiacono:2013dua}
M.~Archidiacono and S.~Hannestad, {\it {Updated constraints on non-standard
  neutrino interactions from Planck}},  {\em JCAP} {\bf 07} (2014) 046,
  [\href{http://www.arxiv.org/abs/1311.3873}{{\tt 1311.3873}}].

\bibitem{Lancaster:2017ksf}
L.~Lancaster, F.-Y. Cyr-Racine, L.~Knox, and Z.~Pan, {\it {A tale of two modes:
  Neutrino free-streaming in the early universe}},  {\em JCAP} {\bf 07} (2017)
  033, [\href{http://www.arxiv.org/abs/1704.06657}{{\tt 1704.06657}}].

\bibitem{Oldengott:2017fhy}
I.~M. Oldengott, T.~Tram, C.~Rampf, and Y.~Y.~Y. Wong, {\it {Interacting
  neutrinos in cosmology: exact description and constraints}},  {\em JCAP} {\bf
  11} (2017) 027, [\href{http://www.arxiv.org/abs/1706.02123}{{\tt
  1706.02123}}].

\bibitem{RoyChoudhury:2020dmd}
S.~Roy~Choudhury, S.~Hannestad, and T.~Tram, {\it {Updated constraints on
  massive neutrino self-interactions from cosmology in light of the $H_0$
  tension}},  {\em JCAP} {\bf 03} (2021) 084,
  [\href{http://www.arxiv.org/abs/2012.07519}{{\tt 2012.07519}}].

\bibitem{Huang:2021dba}
G.-y. Huang and W.~Rodejohann, {\it {Solving the Hubble tension without
  spoiling Big Bang Nucleosynthesis}},  {\em Phys. Rev. D} {\bf 103} (2021)
  123007, [\href{http://www.arxiv.org/abs/2102.04280}{{\tt 2102.04280}}].

\bibitem{EUCLID:2011zbd}
{\bf EUCLID} {\bf Collaboration}, R.~Laureijs {\em et~al.}, {\it {Euclid
  Definition Study Report}},  \href{http://www.arxiv.org/abs/1110.3193}{{\tt
  1110.3193}}.

\bibitem{Maartens:2015mra}
{\bf SKA Cosmology SWG} {\bf Collaboration}, R.~Maartens, F.~B. Abdalla,
  M.~Jarvis, and M.~G. Santos, {\it {Overview of Cosmology with the SKA}},
  {\em PoS} {\bf AASKA14} (2015) 016,
  [\href{http://www.arxiv.org/abs/1501.04076}{{\tt 1501.04076}}].

\bibitem{DESI:2016fyo}
{\bf DESI} {\bf Collaboration}, A.~Aghamousa {\em et~al.}, {\it {The DESI
  Experiment Part I: Science,Targeting, and Survey Design}},
  \href{http://www.arxiv.org/abs/1611.00036}{{\tt 1611.00036}}.

\bibitem{CMB-S4:2016ple}
{\bf CMB-S4} {\bf Collaboration}, K.~N. Abazajian {\em et~al.}, {\it {CMB-S4
  Science Book, First Edition}},
  \href{http://www.arxiv.org/abs/1610.02743}{{\tt 1610.02743}}.

\bibitem{LSSTDarkEnergyScience:2018jkl}
{\bf LSST Dark Energy Science} {\bf Collaboration}, D.~Alonso {\em et~al.},
  {\it {The LSST Dark Energy Science Collaboration (DESC) Science Requirements
  Document}},  \href{http://www.arxiv.org/abs/1809.01669}{{\tt 1809.01669}}.

\bibitem{Blinov:2019gcj}
N.~Blinov, K.~J. Kelly, G.~Z. Krnjaic, and S.~D. McDermott, {\it {Constraining
  the Self-Interacting Neutrino Interpretation of the Hubble Tension}},  {\em
  Phys. Rev. Lett.} {\bf 123} (2019), no.~19 191102,
  [\href{http://www.arxiv.org/abs/1905.02727}{{\tt 1905.02727}}].

\bibitem{Hooper:2007jr}
D.~Hooper, {\it {Detecting MeV Gauge Bosons with High-Energy Neutrino
  Telescopes}},  {\em Phys. Rev. D} {\bf 75} (2007) 123001,
  [\href{http://www.arxiv.org/abs/hep-ph/0701194}{{\tt hep-ph/0701194}}].

\bibitem{Ng:2014pca}
K.~C.~Y. Ng and J.~F. Beacom, {\it {Cosmic neutrino cascades from secret
  neutrino interactions}},  {\em Phys. Rev. D} {\bf 90} (2014), no.~6 065035,
  [\href{http://www.arxiv.org/abs/1404.2288}{{\tt 1404.2288}}]. [Erratum:
  Phys.Rev.D 90, 089904 (2014)].

\bibitem{Ioka:2014kca}
K.~Ioka and K.~Murase, {\it {IceCube PeV\textendash{}EeV neutrinos and secret
  interactions of neutrinos}},  {\em PTEP} {\bf 2014} (2014), no.~6 061E01,
  [\href{http://www.arxiv.org/abs/1404.2279}{{\tt 1404.2279}}].

\bibitem{Kolb:1987qy}
E.~W. Kolb and M.~S. Turner, {\it {Supernova SN 1987a and the Secret
  Interactions of Neutrinos}},  {\em Phys. Rev. D} {\bf 36} (1987) 2895.

\bibitem{Esteban:2020cvm}
I.~Esteban, M.~C. Gonzalez-Garcia, M.~Maltoni, T.~Schwetz, and A.~Zhou, {\it
  {The fate of hints: updated global analysis of three-flavor neutrino
  oscillations}},  {\em JHEP} {\bf 09} (2020) 178,
  [\href{http://www.arxiv.org/abs/2007.14792}{{\tt 2007.14792}}].

\bibitem{deSalas:2020pgw}
P.~F. de~Salas, D.~V. Forero, S.~Gariazzo, P.~Mart\'\i{}nez-Mirav\'e, O.~Mena,
  C.~A. Ternes, M.~T\'ortola, and J.~W.~F. Valle, {\it {2020 global
  reassessment of the neutrino oscillation picture}},  {\em JHEP} {\bf 02}
  (2021) 071, [\href{http://www.arxiv.org/abs/2006.11237}{{\tt 2006.11237}}].

\bibitem{Capozzi:2021fjo}
F.~Capozzi, E.~Di~Valentino, E.~Lisi, A.~Marrone, A.~Melchiorri, and
  A.~Palazzo, {\it {The unfinished fabric of the three neutrino paradigm}},
  \href{http://www.arxiv.org/abs/2107.00532}{{\tt 2107.00532}}.

\bibitem{IceCube:2020acn}
{\bf IceCube} {\bf Collaboration}, M.~G. Aartsen {\em et~al.}, {\it
  {Characteristics of the diffuse astrophysical electron and tau neutrino flux
  with six years of IceCube high energy cascade data}},  {\em Phys. Rev. Lett.}
  {\bf 125} (2020), no.~12 121104,
  [\href{http://www.arxiv.org/abs/2001.09520}{{\tt 2001.09520}}].

\bibitem{IceCube:2020wum}
{\bf IceCube} {\bf Collaboration}, R.~Abbasi {\em et~al.}, {\it {The IceCube
  high-energy starting event sample: Description and flux characterization with
  7.5 years of data}},  \href{http://www.arxiv.org/abs/2011.03545}{{\tt
  2011.03545}}.

\bibitem{IceCube-Gen2:2020qha}
{\bf IceCube-Gen2} {\bf Collaboration}, M.~G. Aartsen {\em et~al.}, {\it
  {IceCube-Gen2: the window to the extreme Universe}},  {\em J. Phys. G} {\bf
  48} (2021), no.~6 060501, [\href{http://www.arxiv.org/abs/2008.04323}{{\tt
  2008.04323}}].

\bibitem{Laha:2013xua}
R.~Laha, B.~Dasgupta, and J.~F. Beacom, {\it {Constraints on New Neutrino
  Interactions via Light Abelian Vector Bosons}},  {\em Phys. Rev. D} {\bf 89}
  (2014), no.~9 093025, [\href{http://www.arxiv.org/abs/1304.3460}{{\tt
  1304.3460}}].

\bibitem{Karshenboim:2014tka}
S.~G. Karshenboim, D.~McKeen, and M.~Pospelov, {\it {Constraints on
  muon-specific dark forces}},  {\em Phys. Rev. D} {\bf 90} (2014), no.~7
  073004, [\href{http://www.arxiv.org/abs/1401.6154}{{\tt 1401.6154}}].
  [Addendum: Phys.Rev.D 90, 079905 (2014)].

\bibitem{Huang:2017egl}
G.-y. Huang, T.~Ohlsson, and S.~Zhou, {\it {Observational Constraints on Secret
  Neutrino Interactions from Big Bang Nucleosynthesis}},  {\em Phys. Rev. D}
  {\bf 97} (2018), no.~7 075009,
  [\href{http://www.arxiv.org/abs/1712.04792}{{\tt 1712.04792}}].

\bibitem{Blum:2014ewa}
K.~Blum, A.~Hook, and K.~Murase, {\it {High energy neutrino telescopes as a
  probe of the neutrino mass mechanism}},
  \href{http://www.arxiv.org/abs/1408.3799}{{\tt 1408.3799}}.

\bibitem{Berryman:2018ogk}
J.~M. Berryman, A.~De~Gouv\^ea, K.~J. Kelly, and Y.~Zhang, {\it
  {Lepton-Number-Charged Scalars and Neutrino Beamstrahlung}},  {\em Phys. Rev.
  D} {\bf 97} (2018), no.~7 075030,
  [\href{http://www.arxiv.org/abs/1802.00009}{{\tt 1802.00009}}].

\bibitem{Kelly:2020pcy}
K.~J. Kelly, M.~Sen, W.~Tangarife, and Y.~Zhang, {\it {Origin of sterile
  neutrino dark matter via secret neutrino interactions with vector bosons}},
  {\em Phys. Rev. D} {\bf 101} (2020), no.~11 115031,
  [\href{http://www.arxiv.org/abs/2005.03681}{{\tt 2005.03681}}].

\bibitem{Brdar:2020nbj}
V.~Brdar, M.~Lindner, S.~Vogl, and X.-J. Xu, {\it {Revisiting neutrino
  self-interaction constraints from $Z$ and $\tau$ decays}},  {\em Phys. Rev.
  D} {\bf 101} (2020), no.~11 115001,
  [\href{http://www.arxiv.org/abs/2003.05339}{{\tt 2003.05339}}].

\bibitem{Kelly:2019wow}
K.~J. Kelly and Y.~Zhang, {\it {Mononeutrino at DUNE: New Signals from
  Neutrinophilic Thermal Dark Matter}},  {\em Phys. Rev. D} {\bf 99} (2019),
  no.~5 055034, [\href{http://www.arxiv.org/abs/1901.01259}{{\tt 1901.01259}}].

\bibitem{Arbey:2011nf}
A.~Arbey, {\it {AlterBBN: A program for calculating the BBN abundances of the
  elements in alternative cosmologies}},  {\em Comput. Phys. Commun.} {\bf 183}
  (2012) 1822--1831, [\href{http://www.arxiv.org/abs/1106.1363}{{\tt
  1106.1363}}].

\bibitem{Arbey:2018zfh}
A.~Arbey, J.~Auffinger, K.~P. Hickerson, and E.~S. Jenssen, {\it {AlterBBN v2:
  A public code for calculating Big-Bang nucleosynthesis constraints in
  alternative cosmologies}},  {\em Comput. Phys. Commun.} {\bf 248} (2020)
  106982, [\href{http://www.arxiv.org/abs/1806.11095}{{\tt 1806.11095}}].

\bibitem{Weiler1982}
T.~Weiler, {\it Resonant absorption of cosmic-ray neutrinos by the
  relic-neutrino background},  {\em Phys. Rev. Lett.} {\bf 49} (Jul, 1982)
  234--237.

\bibitem{Roulet1993}
E.~Roulet, {\it Ultrahigh energy neutrino absorption by neutrino dark matter},
  {\em Phys. Rev. D} {\bf 47} (Jun, 1993) 5247--5252.

\bibitem{Nussinov:1976uw}
S.~Nussinov, {\it {Solar Neutrinos and Neutrino Mixing}},  {\em Phys. Lett. B}
  {\bf 63} (1976) 201--203.

\bibitem{Creque-Sarbinowski:2020qhz}
C.~Creque-Sarbinowski, J.~Hyde, and M.~Kamionkowski, {\it {Resonant neutrino
  self-interactions}},  {\em Phys. Rev. D} {\bf 103} (2021), no.~2 023527,
  [\href{http://www.arxiv.org/abs/2005.05332}{{\tt 2005.05332}}].

\bibitem{Hopkins:2006bw}
A.~M. Hopkins and J.~F. Beacom, {\it {On the normalisation of the cosmic star
  formation history}},  {\em Astrophys. J.} {\bf 651} (2006) 142--154,
  [\href{http://www.arxiv.org/abs/astro-ph/0601463}{{\tt astro-ph/0601463}}].

\bibitem{Yuksel:2008cu}
H.~Yuksel, M.~D. Kistler, J.~F. Beacom, and A.~M. Hopkins, {\it {Revealing the
  High-Redshift Star Formation Rate with Gamma-Ray Bursts}},  {\em Astrophys.
  J. Lett.} {\bf 683} (2008) L5--L8,
  [\href{http://www.arxiv.org/abs/0804.4008}{{\tt 0804.4008}}].

\bibitem{DiFranzo:2015qea}
A.~DiFranzo and D.~Hooper, {\it {Searching for MeV-Scale Gauge Bosons with
  IceCube}},  {\em Phys. Rev. D} {\bf 92} (2015), no.~9 095007,
  [\href{http://www.arxiv.org/abs/1507.03015}{{\tt 1507.03015}}].

\bibitem{Lunardini:2000fy}
C.~Lunardini and A.~Y. Smirnov, {\it {High-energy neutrino conversion and the
  lepton asymmetry in the universe}},  {\em Phys. Rev. D} {\bf 64} (2001)
  073006, [\href{http://www.arxiv.org/abs/hep-ph/0012056}{{\tt
  hep-ph/0012056}}].

\bibitem{Arguelles:2015dca}
C.~A. Arg\"uelles, T.~Katori, and J.~Salvado, {\it {New Physics in
  Astrophysical Neutrino Flavor}},  {\em Phys. Rev. Lett.} {\bf 115} (2015)
  161303, [\href{http://www.arxiv.org/abs/1506.02043}{{\tt 1506.02043}}].

\bibitem{Bustamante:2015waa}
M.~Bustamante, J.~F. Beacom, and W.~Winter, {\it {Theoretically palatable
  flavor combinations of astrophysical neutrinos}},  {\em Phys. Rev. Lett.}
  {\bf 115} (2015), no.~16 161302,
  [\href{http://www.arxiv.org/abs/1506.02645}{{\tt 1506.02645}}].

\bibitem{Ge:2018uhz}
S.-F. Ge and S.~J. Parke, {\it {Scalar Nonstandard Interactions in Neutrino
  Oscillation}},  {\em Phys. Rev. Lett.} {\bf 122} (2019), no.~21 211801,
  [\href{http://www.arxiv.org/abs/1812.08376}{{\tt 1812.08376}}].

\bibitem{Cherry:2016jol}
J.~F. Cherry, A.~Friedland, and I.~M. Shoemaker, {\it {Short-baseline neutrino
  oscillations, Planck, and IceCube}},
  \href{http://www.arxiv.org/abs/1605.06506}{{\tt 1605.06506}}.

\bibitem{DiValentino:2019dzu}
E.~Di~Valentino, A.~Melchiorri, and J.~Silk, {\it {Cosmological constraints in
  extended parameter space from the Planck 2018 Legacy release}},  {\em JCAP}
  {\bf 01} (2020) 013, [\href{http://www.arxiv.org/abs/1908.01391}{{\tt
  1908.01391}}].

\bibitem{RoyChoudhury:2019hls}
S.~Roy~Choudhury and S.~Hannestad, {\it {Updated results on neutrino mass and
  mass hierarchy from cosmology with Planck 2018 likelihoods}},  {\em JCAP}
  {\bf 07} (2020) 037, [\href{http://www.arxiv.org/abs/1907.12598}{{\tt
  1907.12598}}].

\bibitem{Escudero:2020ped}
M.~Escudero, J.~Lopez-Pavon, N.~Rius, and S.~Sandner, {\it {Relaxing
  Cosmological Neutrino Mass Bounds with Unstable Neutrinos}},  {\em JHEP} {\bf
  12} (2020) 119, [\href{http://www.arxiv.org/abs/2007.04994}{{\tt
  2007.04994}}].

\bibitem{Esteban:2021ozz}
I.~Esteban and J.~Salvado, {\it {Long Range Interactions in Cosmology:
  Implications for Neutrinos}},  {\em JCAP} {\bf 05} (2021) 036,
  [\href{http://www.arxiv.org/abs/2101.05804}{{\tt 2101.05804}}].

\bibitem{Beacom:2002cb}
J.~F. Beacom and N.~F. Bell, {\it {Do Solar Neutrinos Decay?}},  {\em Phys.
  Rev. D} {\bf 65} (2002) 113009,
  [\href{http://www.arxiv.org/abs/hep-ph/0204111}{{\tt hep-ph/0204111}}].

\bibitem{Zralek:1997sa}
M.~Zralek, {\it {On the possibilities of distinguishing Dirac from Majorana
  neutrinos}},  {\em Acta Phys. Polon. B} {\bf 28} (1997) 2225--2257,
  [\href{http://www.arxiv.org/abs/hep-ph/9711506}{{\tt hep-ph/9711506}}].

\bibitem{Rasmussen:2017ert}
R.~W. Rasmussen, L.~Lechner, M.~Ackermann, M.~Kowalski, and W.~Winter, {\it
  {Astrophysical neutrinos flavored with Beyond the Standard Model physics}},
  {\em Phys. Rev. D} {\bf 96} (2017), no.~8 083018,
  [\href{http://www.arxiv.org/abs/1707.07684}{{\tt 1707.07684}}].

\bibitem{Song:2020nfh}
N.~Song, S.~W. Li, C.~A. Arg\"uelles, M.~Bustamante, and A.~C. Vincent, {\it
  {The Future of High-Energy Astrophysical Neutrino Flavor Measurements}},
  {\em JCAP} {\bf 04} (2021) 054,
  [\href{http://www.arxiv.org/abs/2012.12893}{{\tt 2012.12893}}].

\bibitem{Bustamante:2020mep}
M.~Bustamante, C.~Rosenstr\o{}m, S.~Shalgar, and I.~Tamborra, {\it {Bounds on
  secret neutrino interactions from high-energy astrophysical neutrinos}},
  {\em Phys. Rev. D} {\bf 101} (2020), no.~12 123024,
  [\href{http://www.arxiv.org/abs/2001.04994}{{\tt 2001.04994}}].

\bibitem{Mazumdar:2020ibx}
A.~Mazumdar, S.~Mohanty, and P.~Parashari, {\it {Flavour specific neutrino
  self-interaction: $H_0$ tension and IceCube}},
  \href{http://www.arxiv.org/abs/2011.13685}{{\tt 2011.13685}}.

\bibitem{Laha:2013lka}
R.~Laha, J.~F. Beacom, B.~Dasgupta, S.~Horiuchi, and K.~Murase, {\it
  {Demystifying the PeV Cascades in IceCube: Less (Energy) is More (Events)}},
  {\em Phys. Rev. D} {\bf 88} (2013) 043009,
  [\href{http://www.arxiv.org/abs/1306.2309}{{\tt 1306.2309}}].

\bibitem{Glashow:1960zz}
S.~L. Glashow, {\it {Resonant Scattering of Antineutrinos}},  {\em Phys. Rev.}
  {\bf 118} (1960) 316--317.

\bibitem{IceCube:2021rpz}
{\bf IceCube} {\bf Collaboration}, M.~G. Aartsen {\em et~al.}, {\it {Detection
  of a particle shower at the Glashow resonance with IceCube}},  {\em Nature}
  {\bf 591} (2021), no.~7849 220--224. [Erratum: Nature 592, E11 (2021)].

\bibitem{Abbasi:2020jmh}
{\bf IceCube} {\bf Collaboration}, R.~Abbasi {\em et~al.}, {\it {The IceCube
  high-energy starting event sample: Description and flux characterization with
  7.5 years of data}},  \href{http://www.arxiv.org/abs/2011.03545}{{\tt
  2011.03545}}.

\bibitem{Fermi:1949ee}
E.~Fermi, {\it {On the Origin of the Cosmic Radiation}},  {\em Phys. Rev.} {\bf
  75} (1949) 1169--1174.

\bibitem{Gaisser:2016uoy}
T.~K. Gaisser, R.~Engel, and E.~Resconi, {\em {Cosmic Rays and Particle
  Physics}: {2nd Edition}}.
\newblock Cambridge University Press, 6, 2016.

\bibitem{Aartsen:2015zva}
{\bf IceCube} {\bf Collaboration}, M.~G. Aartsen {\em et~al.}, {\it {The
  IceCube Neutrino Observatory - Contributions to ICRC 2015 Part II:
  Atmospheric and Astrophysical Diffuse Neutrino Searches of All Flavors}},  in
  {\em {34th International Cosmic Ray Conference}}, 10, 2015.
\newblock \href{http://www.arxiv.org/abs/1510.05223}{{\tt 1510.05223}}.

\bibitem{Palladino:2017qda}
A.~Palladino, C.~Mascaretti, and F.~Vissani, {\it {On the compatibility of the
  IceCube results with a universal neutrino spectrum}},  {\em Eur. Phys. J. C}
  {\bf 77} (2017), no.~10 684, [\href{http://www.arxiv.org/abs/1708.02094}{{\tt
  1708.02094}}].

\bibitem{KM3Net:2016zxf}
{\bf KM3Net} {\bf Collaboration}, S.~Adrian-Martinez {\em et~al.}, {\it {Letter
  of intent for KM3NeT 2.0}},  {\em J. Phys. G} {\bf 43} (2016), no.~8 084001,
  [\href{http://www.arxiv.org/abs/1601.07459}{{\tt 1601.07459}}].

\bibitem{Baikal-GVD:2018isr}
{\bf Baikal-GVD} {\bf Collaboration}, A.~D. Avrorin {\em et~al.}, {\it
  {Baikal-GVD: status and prospects}},  {\em EPJ Web Conf.} {\bf 191} (2018)
  01006, [\href{http://www.arxiv.org/abs/1808.10353}{{\tt 1808.10353}}].

\bibitem{P-ONE:2020ljt}
{\bf P-ONE} {\bf Collaboration}, M.~Agostini {\em et~al.}, {\it {The Pacific
  Ocean Neutrino Experiment}},  {\em Nature Astron.} {\bf 4} (2020), no.~10
  913--915, [\href{http://www.arxiv.org/abs/2005.09493}{{\tt 2005.09493}}].

\bibitem{Gandhi:1998ri}
R.~Gandhi, C.~Quigg, M.~H. Reno, and I.~Sarcevic, {\it {Neutrino interactions
  at ultrahigh-energies}},  {\em Phys. Rev. D} {\bf 58} (1998) 093009,
  [\href{http://www.arxiv.org/abs/hep-ph/9807264}{{\tt hep-ph/9807264}}].

\bibitem{thesis}
A.~Schneider, ``Precision measurements of the astro-physical neutrino flux.''
\newblock
  \url{https://raw.githubusercontent.com/austinschneider/gradthesis/master/thesis_submitted.pdf}.

\bibitem{GRAND:2015uko}
{\bf GRAND} {\bf Collaboration}, O.~Martineau-Huynh {\em et~al.}, {\it {The
  Giant Radio Array for Neutrino Detection}},  {\em EPJ Web Conf.} {\bf 116}
  (2016) 03005, [\href{http://www.arxiv.org/abs/1508.01919}{{\tt 1508.01919}}].

\bibitem{ARA:2019wcf}
{\bf ARA} {\bf Collaboration}, P.~Allison {\em et~al.}, {\it {Constraints on
  the diffuse flux of ultrahigh energy neutrinos from four years of Askaryan
  Radio Array data in two stations}},  {\em Phys. Rev. D} {\bf 102} (2020),
  no.~4 043021, [\href{http://www.arxiv.org/abs/1912.00987}{{\tt 1912.00987}}].

\bibitem{Abarr:2020bjd}
Q.~Abarr {\em et~al.}, {\it {The Payload for Ultrahigh Energy Observations
  (PUEO): A White Paper}},  \href{http://www.arxiv.org/abs/2010.02892}{{\tt
  2010.02892}}.

\bibitem{POEMMA:2020ykm}
{\bf POEMMA} {\bf Collaboration}, A.~V. Olinto {\em et~al.}, {\it {The POEMMA
  (Probe of Extreme Multi-Messenger Astrophysics) observatory}},  {\em JCAP}
  {\bf 06} (2021) 007, [\href{http://www.arxiv.org/abs/2012.07945}{{\tt
  2012.07945}}].

\bibitem{Prohira:2021vvn}
S.~Prohira {\em et~al.}, {\it {The Radar Echo Telescope for Cosmic Rays:
  Pathfinder Experiment for a Next-Generation Neutrino Observatory}},
  \href{http://www.arxiv.org/abs/2104.00459}{{\tt 2104.00459}}.

\bibitem{Fiorillo:2020jvy}
D.~F.~G. Fiorillo, G.~Miele, S.~Morisi, and N.~Saviano, {\it {Cosmogenic
  neutrino fluxes under the effect of active-sterile secret interactions}},
  {\em Phys. Rev. D} {\bf 101} (2020), no.~8 083024,
  [\href{http://www.arxiv.org/abs/2002.10125}{{\tt 2002.10125}}].

\bibitem{Fiorillo:2020zzj}
D.~F.~G. Fiorillo, S.~Morisi, G.~Miele, and N.~Saviano, {\it {Observable
  features in ultrahigh energy neutrinos due to active-sterile secret
  interactions}},  {\em Phys. Rev. D} {\bf 102} (2020), no.~8 083014,
  [\href{http://www.arxiv.org/abs/2007.07866}{{\tt 2007.07866}}].

\bibitem{Kelly:2018tyg}
K.~J. Kelly and P.~A.~N. Machado, {\it {Multimessenger Astronomy and New
  Neutrino Physics}},  {\em JCAP} {\bf 10} (2018) 048,
  [\href{http://www.arxiv.org/abs/1808.02889}{{\tt 1808.02889}}].

\bibitem{Haber:1994pe}
H.~E. Haber, {\it {Spin formalism and applications to new physics searches}},
  in {\em {21st Annual SLAC Summer Institute on Particle Physics: Spin
  Structure in High-energy Processes (School: 26 Jul - 3 Aug, Topical
  Conference: 4-6 Aug) (SSI 93)}}, pp.~231--272, 4, 1994.
\newblock \href{http://www.arxiv.org/abs/hep-ph/9405376}{{\tt hep-ph/9405376}}.

\bibitem{Zyla:2020zbs}
{\bf Particle Data Group} {\bf Collaboration}, P.~Zyla {\em et~al.}, {\it
  {Review of Particle Physics}},  {\em PTEP} {\bf 2020} (2020), no.~8 083C01.

\bibitem{Denner:1992me}
A.~Denner, H.~Eck, O.~Hahn, and J.~Kublbeck, {\it {Compact Feynman rules for
  Majorana fermions}},  {\em Phys. Lett. B} {\bf 291} (1992) 278--280.

\end{thebibliography}\endgroup

\clearpage
\onecolumngrid
\appendix
\section{Cross-Section Formulae}
\label{sec:appendix}

\begin{figure}[hbtp]
\subfloat[]{\label{fig:Feynman_nu_a}
    \includegraphics[width=0.24\textwidth,valign=c]{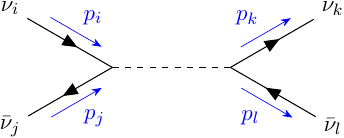}
}\hspace{0.01\textwidth}
\subfloat[]{\label{fig:Feynman_nu_b}    \includegraphics[height=0.14\textheight,valign=c]{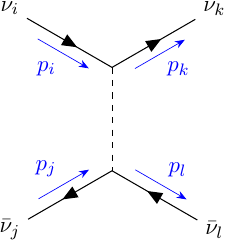}
}\hspace{0.01\textwidth}  \subfloat[]{\label{fig:Feynman_nu_c}
    \includegraphics[height=0.14\textheight,valign=c]{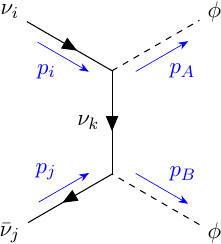}
}\hspace{0.01\textwidth}
  \subfloat[]{\label{fig:Feynman_nu_d}
    \includegraphics[height=0.14\textheight,valign=c]{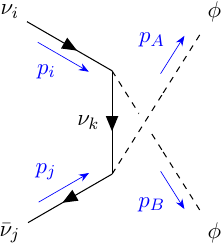}
}

  \subfloat[]{\label{fig:Feynman_nu_e}
    \includegraphics[height=0.14\textheight,valign=c]{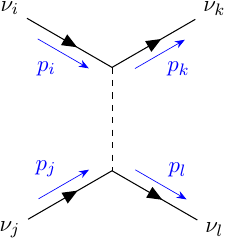}
}\hspace{0.01\textwidth}
\subfloat[]{\label{fig:Feynman_nu_f}    \includegraphics[height=0.14\textheight,valign=c]{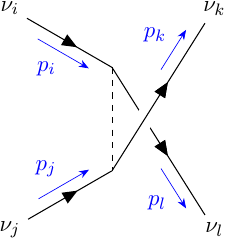}
}
  \caption{Relevant Feynman diagrams for neutrino
    scattering at lowest order.}
  \label{fig:Feynman_nu}  
\end{figure}

In this Appendix, we derive the full neutrino-neutrino scattering
cross-section necessary for the calculations in the main text. The
relevant Feynman diagrams stemming from the Lagrangian~\eqref{eq:Lag} are shown in 
\cref{fig:Feynman_nu}:
\cref{fig:Feynman_nu_a,fig:Feynman_nu_b,fig:Feynman_nu_c,fig:Feynman_nu_d}
correspond to neutrino-antineutrino scattering, and
\cref{fig:Feynman_nu_e,fig:Feynman_nu_f} to neutrino-neutrino
scattering, where $p$ denote momenta and lowercase subindices denote mass
eigenstates.

The amplitudes for the Feynman diagrams in \cref{fig:Feynman_nu} are

\noindent
\begin{tabular}{@{}p{.5\linewidth}@{}p{.5\linewidth}@{}}
  {\begin{align}
    i \mathcal{M}_a &= -i g_{ij}^* g_{kl} \, \frac{\bar{\mathrm{u}}_k \mathrm{v}_l \bar{\mathrm{v}}_j \mathrm{u}_i}{s
      - M_\phi^2 + i M_\phi \Gamma} \, , \\
i \mathcal{M}_b &= i g_{ik}^* g_{jl} \, \frac{\bar{\mathrm{u}}_k \mathrm{u}_i \bar{\mathrm{v}}_l \mathrm{v}_j}{t
  - M_\phi^2} \, , \\
i \mathcal{M}_c &= -i \sum_k g_{ik}^* g_{jk} \, \bar{\mathrm{v}}_j
\frac{\slashed{p}_i - \slashed{p}_A + m_k}{t - m_k^2}\mathrm{u}_i \, ,
  \end{align}} &   {\begin{align}
    i \mathcal{M}_d &= -i \sum_k g_{ik}^* g_{jk} \, \bar{\mathrm{v}}_j
    \frac{\slashed{p}_i - \slashed{p}_B + m_k}{u - m_k^2}\mathrm{u}_i \, , \\
    i \mathcal{M}_e &= i g_{ik}^* g_{jl}^* \, \frac{\bar{\mathrm{u}}_k \mathrm{u}_i \bar{\mathrm{u}}_l \mathrm{u}_j}{t
      - M_\phi^2} \, , \\
    i \mathcal{M}_f &= -i g_{il}^* g_{jk}^* \, \frac{\bar{\mathrm{u}}_k \mathrm{u}_j \bar{\mathrm{u}}_l \mathrm{u}_i}{u
  - M_\phi^2} \, , 
  \end{align}}
\end{tabular}
where $g_{ij} = \sum_{\alpha, \, \beta} U^*_{\alpha i} U_{\beta j} g_{\alpha \beta}$ is the coupling strength in the mass basis, $\mathrm{u}$ and $\mathrm{v}$ are the standard Dirac spinors, $s$, $t$ and $u$ are
the Mandelstam variables,\footnote{We define, for a neutrino final state, $t
\equiv (p_i - p_k)^2$; and for a scalar final state, $t \equiv (p_i -
p_A)^2$.} $m_k$ is the mass of the k-th neutrino mass eigenstate,
$M_\phi$ is the scalar mass, and $\Gamma$ is the scalar decay width.

Our working assumptions are:
\begin{enumerate}
\item The incident neutrino is ultrarelativistic and left-handed (see
  Ref.~\cite{Haber:1994pe} for the polarized scattering formalism). 
\item We neglect CP violation (i.e., complex phases) in
  $g_{ij}$. One can check that if the scalar couples to a single
  neutrino flavor, CP-violating effects are not present in
  neutrino-neutrino scattering.
\item The target neutrino is at rest and unpolarized.
\end{enumerate}
Under these assumptions, the differential scattering
cross-section is given by~\cite{Zyla:2020zbs}
\begin{equation}
\frac{\mathrm{d}\sigma}{\mathrm{d}t} = \left(\frac{1}{2}\right)
\frac{1}{16 \pi s^2} |\mathcal{M}|^2 \, ,
\end{equation}
where the 1/2 factor applies for identical final-state
particles. The kinematic limits for two neutrinos in the final state
are ${t \in [-s, 0]}$, whereas for double-scalar production, ${t \in \left[-
  \left( \sqrt{s}/2 + \sqrt{s/4-M_\phi^2}\right), -
  \left( \sqrt{s}/2 -
  \sqrt{s/4-M_\phi^2}\right)\right]}$.

\subsection{Dirac neutrinos}

If neutrinos are Dirac fermions, and assuming that there is the same amount of
target neutrinos and antineutrinos, the total inclusive cross-section
for initial mass eigenstate $i$ and final mass eigenstate $j$ is given
by
\begin{equation}
\begin{split}
\sigma_{ij}(s) = \frac{1}{16\pi} \sum_{k, \,l} &\left[ g_{ij}^2g_{kl}^2 \frac{s}{(s-M_\phi^2)^2+M_\phi^2\Gamma^2} \right. + \left(\frac{3}{2} \, g_{ik}^2 g_{jl}^2 + \frac{1}{2} \, g_{il}^2 g_{jk}^2\right)\left(\frac{s+2M_\phi^2}{s(s+M_\phi^2)} + 2 \frac{M_\phi^2}{s^2} \log \frac{M_\phi^2}{M_\phi^2 + s}\right) \\
&- g_{ij} g_{kl} g_{ik} g_{jl}  \left(\frac{M_\phi^2}{s}-1\right)
  \frac{s+M_\phi^2\log \frac{M_\phi^2}{M_\phi^2 +
      s}}{(s-M_\phi^2)^2+M_\phi^2\Gamma^2}  \left.+ \frac{1}{2} \,
  g_{ik} g_{jl} g_{il} g_{jk} \left(\frac{1}{s} +
  \frac{2M_\phi^2(M_\phi^2+s)}{s^2(2M_\phi^2+s)} \log
  \frac{M_\phi^2}{M_\phi^2 + s} \right) \right] \\
+ \frac{1}{64 \pi s^2} & \left(\sum_k g_{ik}^2 g_{jk}^2\right) \left\lbrace\frac{s^2-4 M_\phi^2 s+6 M_\phi^4}{s-2M_\phi^2} \log \left[ \left(\frac{\sqrt{s(s-4M_\phi^2)}+s-2M_\phi^2}{\sqrt{s(s-4M_\phi^2)}-s+2M_\phi^2}\right)^2 \right]-6\sqrt{s(s-4 M_\phi^2)}  \vphantom{\log \left[ \left(\frac{\sqrt{s(s-4M_\phi^2)}+s-2M_\phi^2}{\sqrt{s(s-4M_\phi^2)}-s+2M_\phi^2}\right)^2 \right]} \right\rbrace\, .
\end{split}
\end{equation}
The first sum is the cross-section with two neutrinos in the final
state, and the last term is the cross-section for
double-scalar production, present only if $s > 4 M_\phi^2$.

For double-neutrino production, we can classify the differential
cross-section in terms of the visibility of the final state. Here,
visible means that neutrinos must be left-handed and antineutrinos
right-handed. We show the results in \cref{tab:dsigma}.

\begin{table}[h]
\begin{adjustbox}{center}
\begin{tabular}{c|c}
& $\displaystyle \frac{\mathrm{d}\sigma}{\mathrm{d}t}$ \\ \\ \hline \\
$k$ visible, $l$ invisible & $\displaystyle \frac{1}{32\pi} \left[ g_{ij}^2g_{kl}^2 \frac{1}{(s-M_\phi^2)^2+M_\phi^2\Gamma^2} + \left( \frac{1}{2} \right) g_{il}^2 g_{jk}^2 \frac{u^2/s^2}{(u-M_\phi^2)^2}\right]$ \\[1cm]
$k$ invisible, $l$ invisible & $\displaystyle \frac{1}{32\pi} \left[ \mathcal{S}_t \, g_{ik}^2 g_{jl}^2 \frac{t^2/s^2}{(t-M_\phi^2)^2} +  \left( \frac{1}{2} \right)\left(g_{il}^2 g_{jk}^2 \frac{u^2/s^2}{(u-M_\phi^2)^2} 
  + 2 g_{ik} g_{jl} g_{il} g_{jk} \frac{t \cdot u/s^2}{(u-M_\phi^2)(t-M_\phi^2)}\right)\right]$\\[1cm]
$k$ invisible, $l$ visible & $\displaystyle \frac{1}{32\pi} \left[ g_{ij}^2 g_{kl}^2 \frac{1}{(s-M_\phi^2)^2+M_\phi^2\Gamma^2} + \mathcal{S}_t \, g_{ik}^2 g_{jl}^2 \frac{t^2/s^2}{(t-M_\phi^2)^2}
  + 2 g_{ij} g_{kl} g_{ik} g_{jl} \frac{t(1-M_\phi^2/s)}{[(s-M_\phi^2)^2+M_\phi^2\Gamma^2](t-M_\phi^2)} \right]$\\[1cm]
$k$ visible, $l$ visible & $0$ \\
\end{tabular}
\end{adjustbox}
\caption{Total differential cross-section for double-neutrino
  production, classified by the final state.}
\label{tab:dsigma}
\end{table}
The 1/2 factors are present if $k=l$, and
\begin{equation}
\mathcal{S}_t \equiv \begin{cases}
\frac{3}{2} \, & \mathrm{if } \, k \neq l \\
2 \, & \mathrm{if } \, k = l
\end{cases} \, .
\end{equation}

For double-scalar production, present only if $s>4 M_\phi^2$, the differential cross-section is given
by
\begin{equation}
\frac{\mathrm{d}\sigma}{\mathrm{d}t} = -\frac{1}{64 \pi s^2} \left(\sum_k g_{ik}^2 g_{jk}^2\right) \frac{(-2M_\phi^2+s+2t)^2}{t^2(-2M_\phi^2+s+t)^2}\left[s t+(t-M_\phi^2)^2\right] \, .
\end{equation}

\subsection{Majorana neutrinos}

If neutrinos are Majorana fermions, there is no distinction between neutrinos and
antineutrinos. Following Ref.~\cite{Denner:1992me}, we obtain the
total inclusive cross-section
for initial mass eigenstate $i$ and final mass eigenstate $j$ as
\begin{equation}
\begin{split}
\sigma_{ij}(s) = \frac{1}{16\pi} \sum_{k, \,l} &\left[ g_{ij}^2g_{kl}^2 \frac{s}{(s-M_\phi^2)^2+M_\phi^2\Gamma^2} \right. + \left(\, g_{ik}^2 g_{jl}^2 + \, g_{il}^2 g_{jk}^2\right)\left(\frac{s+2M_\phi^2}{s(s+M_\phi^2)} + 2 \frac{M_\phi^2}{s^2} \log \frac{M_\phi^2}{M_\phi^2 + s}\right) \\
&- g_{ij} g_{kl} (g_{ik} g_{jl} + g_{il} g_{jk})  \left(\frac{M_\phi^2}{s}-1\right)
  \frac{s+M_\phi^2\log \frac{M_\phi^2}{M_\phi^2 +
      s}}{(s-M_\phi^2)^2+M_\phi^2\Gamma^2}   \left.+ \,
  g_{ik} g_{jl} g_{il} g_{jk} \left(\frac{1}{s} +
  \frac{2M_\phi^2(M_\phi^2+s)}{s^2(2M_\phi^2+s)} \log
  \frac{M_\phi^2}{M_\phi^2 + s} \right) \right] \\
+ \frac{1}{32 \pi s^2} & \left(\sum_k g_{ik}^2 g_{jk}^2\right) \left\lbrace\frac{s^2-4 M_\phi^2 s+6 M_\phi^4}{s-2M_\phi^2} \log \left[ \left(\frac{\sqrt{s(s-4M_\phi^2)}+s-2M_\phi^2}{\sqrt{s(s-4M_\phi^2)}-s+2M_\phi^2}\right)^2 \right]-6\sqrt{s(s-4 M_\phi^2)}  \vphantom{\log \left[ \left(\frac{\sqrt{s(s-4M_\phi^2)}+s-2M_\phi^2}{\sqrt{s(s-4M_\phi^2)}-s+2M_\phi^2}\right)^2 \right]} \right\rbrace\, ,
\end{split}
\end{equation}
The first sum is the cross-section with two neutrinos in the final
state, and the last term is the cross-section for
double-scalar production, present only if $s > 4 M_\phi^2$. 

\begin{figure}[hbtp]
    \centering
    \includegraphics[width=0.5\textwidth]{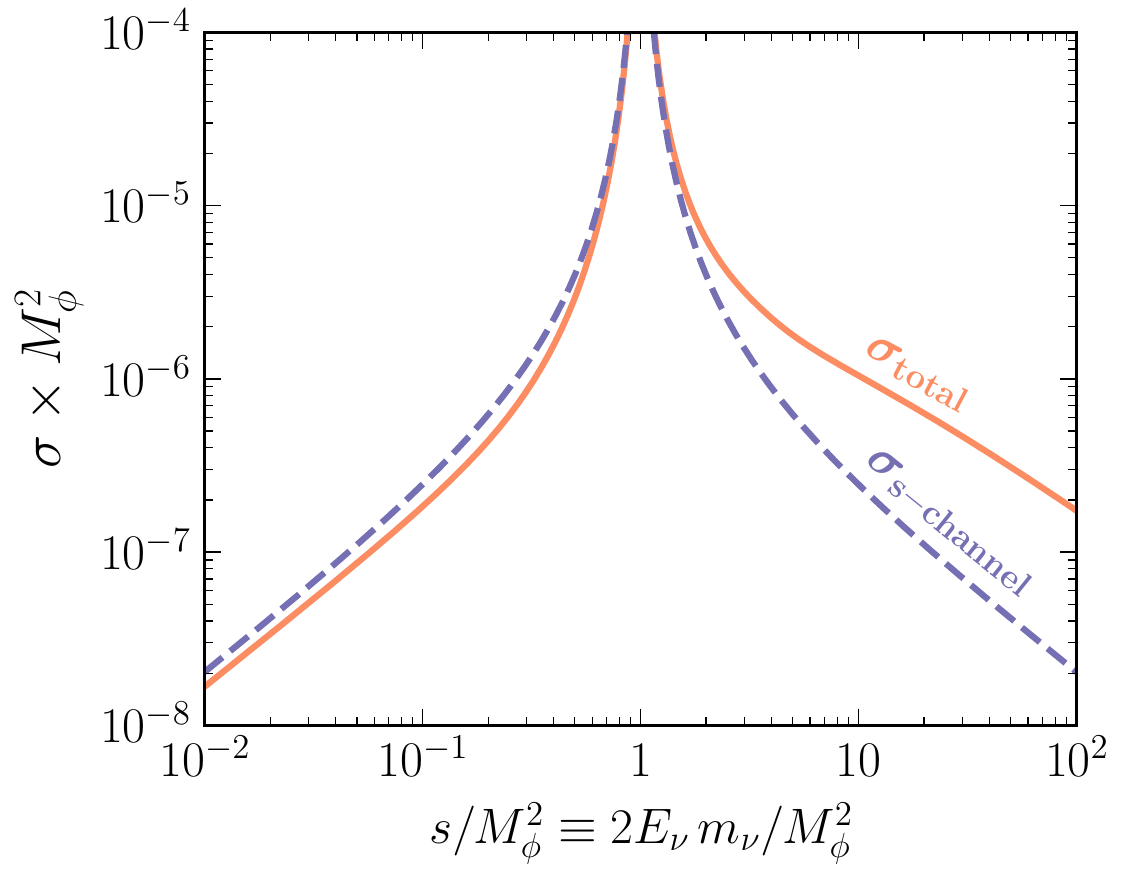}
    \caption{Total interaction cross-section and s-channel contribution. We assume Majorana neutrinos, a single interacting flavor, and a coupling constant $g = 0.1$.}
    \label{fig:xsec_total}
\end{figure}

\Cref{fig:xsec_total} shows the total cross-section as well as the s-channel contribution. As discussed in the main text, for energies above the resonance the s-channel contribution is subdominant. The slight suppression of the total cross-section at lower energies is due to destructive interference between different diagrams.

The differential cross-section is simpler than in the Dirac neutrino case, as
all final states are observable. For double-neutrino production, we
get
\begin{equation}
\begin{split}
\frac{\mathrm{d}\sigma}{\mathrm{d}t} = \left(\frac{1}{2}\right) \frac{1}{8 \pi} & \left[ g_{ij}^2 g_{kl}^2 \frac{1}{(s-M_\phi^2)^2+M_\phi^2\Gamma^2} + g_{ik}^2 g_{jl}^2 \frac{t^2/s^2}{(t-M_\phi^2)^2} 
  + g_{il}^2 g_{jk}^2 \frac{u^2/s^2}{(u-M_\phi^2)^2}  \right. \\
  & + g_{ij} g_{kl} g_{ik} g_{jl} \frac{t(1-M_\phi^2/s)}{[(s-M_\phi^2)^2+M_\phi^2\Gamma^2](t-M_\phi^2)}  + g_{ik} g_{jl} g_{il} g_{jk} \frac{t \cdot u/s^2}{(u-M_\phi^2)(t-M_\phi^2)} \\
  & \left. + g_{ij} g_{kl} g_{il} g_{jk} \frac{u(1-M_\phi^2/s)}{[(s-M_\phi^2)^2+M_\phi^2\Gamma^2](u-M_\phi^2)}\right] \, ,
\end{split}
\end{equation}
where the $1/2$ factor is present if $k=l$.

For double-scalar production, present only if $s > 4 M_\phi^2$, the differential cross-section is given
by
\begin{equation}
\frac{\mathrm{d}\sigma}{\mathrm{d}t} = -\frac{1}{32 \pi s^2} \left(\sum_k g_{ik}^2 g_{jk}^2\right) \frac{(-2M_\phi^2+s+2t)^2}{t^2(-2M_\phi^2+s+t)^2}\left[s t+(t-M_\phi^2)^2\right] \, .
\end{equation}

In all the expressions in this subsection, the cross-section must be
multiplied by \emph{half} the number of targets to obtain the number
of events. In this convention, the number of targets is the same for
Dirac and Majorana neutrinos, assuming the same amount of target
neutrinos and antineutrinos in the Dirac neutrino case.

\subsection{Scalar decay}

\begin{figure}[hbtp]
    \centering
    \includegraphics[width=0.24\textwidth]{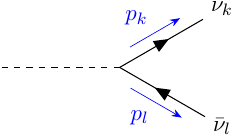}
    \caption{Feynman diagram mediating scalar decay at lowest order.}
    \label{fig:Feynman_decay}
\end{figure}

The only remaining calculation is the scalar decay width. At lowest
order, it is mediated by the Feynman diagram in
\cref{fig:Feynman_decay}.

The amplitude can be immediately computed
\begin{equation}
i \mathcal{M}_\mathrm{decay} = - i g_{kl} \, \bar{\mathrm{u}}_k \mathrm{v}_l \, .
\end{equation}
Summing over final states, we get the total decay
width~\cite{Zyla:2020zbs}
\begin{equation}
  \Gamma = \left(\frac{1}{2}\right) \frac{1}{8\pi}
  \frac{|\vec{p}_k|}{M_\phi^2} |\mathcal{M}|^2 =
  \left(\frac{1}{2}\right) \frac{M_\phi}{8\pi} \sum_{k,\, l} g_{kl}^2
  \, ,
\end{equation}
where the $1/2$ factor is present if neutrinos are Majorana fermions.

For the double-scalar production process (\cref{fig:Feynman_nu_c,fig:Feynman_nu_d}), we also need the
differential decay width
\begin{equation}
\frac{1}{\Gamma} \frac{\mathrm{d}\Gamma}{\mathrm{d}E_k} =
\frac{1}{\Gamma} \frac{\mathrm{d}\Gamma}{\mathrm{d}E_l} =
\frac{1}{E_\phi} \, ,
\end{equation}
where $E_\phi$ is the energy of the scalar. Unlike the total decay
width, this computation is also valid if the scalar is not at rest. If
neutrinos are Dirac fermions, one final state neutrino is visible and the other
is invisible.

\end{document}